\documentclass[10pt,aps,prd,nofootinbib,superscriptaddress,twocolumn,preprintnumbers,balancelastpage]{revtex4}

\usepackage{graphicx,epsfig,psfrag,amssymb,hyperref}
\usepackage{multirow}
\usepackage{color,graphicx,epsfig,psfrag,amsmath,empheq}
\usepackage{bm}
\usepackage{mathrsfs,amsfonts,color}
\usepackage{slashed}
\usepackage{hepunits}
\usepackage{color}
\usepackage{enumitem}
\usepackage{xfrac}

\newcommand{\cC}{\mathcal{C}}
\newcommand{\cF}{\mathcal{F}}

\newcommand{\cL}{\mathcal{L}}

\newcommand{\BR}{{\rm BR}}

\definecolor{greenp1}{rgb}{0, 0.8, 0}
\definecolor{danielColor}{rgb}{0.9, 0.2, 0.9}

\newcommand{\mixa}[1]{\ensuremath{\langle \boldsymbol{a #1} \rangle}\xspace}
\newcommand{\mixatwo}[2]{\ensuremath{\langle \boldsymbol{a #1 #2} \rangle}\xspace}
\newcommand{\chipt}{$\chi$PT\xspace}
\newcommand{\alphaem}{\ensuremath{\alpha_{\rm EM}}\xspace}

\begin{document}

\title{Coupling QCD-scale axion-like particles to gluons}

\author{Daniel Aloni}
\email{daniel.aloni@weizmann.ac.il}
\affiliation{Department of Particle Physics and Astrophysics, Weizmann Institute of Science, Rehovot, Israel 7610001}

\author{Yotam Soreq}
\email{yotam.soreq@cern.ch}
\affiliation{Center for Theoretical Physics, Massachusetts Institute of Technology, Cambridge, MA 02139, U.S.A.}
\affiliation{Theoretical Physics Department, CERN, CH-1211 Geneva 23, Switzerland}

\author{Mike Williams}
\email{mwill@mit.edu}
\affiliation{Laboratory for Nuclear Science, Massachusetts Institute of Technology, Cambridge, MA 02139, U.S.A.}

\begin{abstract}
We present a novel data-driven method for determining the hadronic interaction strengths of axion-like particles~(ALPs) with  QCD-scale  masses.
Using our method, it is possible to calculate the hadronic production and decay rates of ALPs, along with many of the largest ALP decay rates to exclusive final states.
To illustrate the impact on QCD-scale ALP phenomenology, we consider the scenario where the ALP-gluon coupling is dominant over the ALP coupling to photons, electroweak bosons, and all fermions for $m_{\pi} \lesssim m_a \lesssim 3$\,GeV.
We emphasize, however, that our method can easily be generalized to any set of ALP couplings to SM particles.
Finally, using the approach developed here, we provide calculations for the branching fractions of $\eta_c \to VV$ decays, {\em i.e.}\ $\eta_c$ decays into two vector mesons, which are consistent with the known experimental values.
\end{abstract}

\preprint{CERN-TH-2018-237, MIT-CTP/5080}
\maketitle

Axion-like particles~(ALPs) are hypothetical pseudoscalars whose couplings to the gauge bosons of the Standard Model (SM)---the gluons, photons, and electroweak bosons---are highly suppressed at low energies by a large cut-off scale $\Lambda$.
ALPs are found in many proposed extensions to the SM (see Refs.\cite{Essig:2013lka,Marsh:2015xka,Graham:2015ouw,Irastorza:2018dyq}),
since they naturally address such puzzles as the Strong $CP$\,\cite{Peccei:1977hh,Peccei:1977ur,Weinberg:1977ma,Wilczek:1977pj}  and Hierarchy problems~\cite{Graham:2015cka}.
Moreover, ALPs may explain the long-standing anomaly with the magnetic moment of the muon~\cite{Chang:2000ii},
and could provide a {\em portal} connecting SM particles to dark matter~\cite{Nomura:2008ru,Freytsis:2010ne,Dolan:2014ska,Hochberg:2018rjs}.

ALPs are pseudo-Nambu-Goldstone bosons, and therefore, their masses, $m_a$, are expected to be $m_a \ll \Lambda$. 
Recently, MeV-to-GeV scale, henceforth QCD-scale, ALPs have received considerable interest\,\cite{Alves:2017avw,Marciano:2016yhf,Jaeckel:2015jla,Dobrich:2015jyk,Izaguirre:2016dfi,Knapen:2016moh,Artamonov:2009sz,Fukuda:2015ana,Bauer:2018uxu,Mariotti:2017vtv,CidVidal:2018blh};
however, the phenomenological impact of ALP-gluon interactions is not well understood for QCD-scale ALPs.
The effective Lagrangian describing such interactions is
\begin{align}
	\label{eq:L}
	\mathcal{L}
	\supset
	-\frac{4\pi \alpha_s c_{g}}{\Lambda}a G^{\mu\nu}\tilde{G}_{\mu\nu} \, ,
\end{align}
where $c_{g}$ is the dimensionless $agg$ vertex coupling constant
and $\tilde{G}_{\mu\nu} \equiv \frac{1}{2}\epsilon_{\mu\nu\alpha\beta} G^{\alpha\beta}$.

In this Letter, we present a novel data-driven method for determining the hadronic interaction strengths of QCD-scale ALPs.
Using our method, it is possible to calculate the hadronic production and decay rates of ALPs, along with many of the largest ALP decay branching fractions to exclusive final states.
To illustrate the impact on QCD-scale ALP phenomenology of $c_g \neq 0$, 
we consider
\begin{align}
  	\label{eq:cvals}
	c_g \gg c_{\gamma}, c_{\rm EW}, c_f \, ,
\end{align}
for $m_{\pi} \lesssim m_a \lesssim 3$\,GeV; {\em i.e.}\ the scenario where the ALP-gluon coupling is dominant over the ALP coupling to photons~($c_\gamma$), electroweak bosons~($c_{\rm EW}$), and all fermions~($c_{f}$).
We emphasize, however, that our method can easily be generalized to any ALP couplings to SM particles.
The impact of ALP couplings to photons, electroweak bosons, leptons, and heavy quarks is known~\cite{Bauer:2017ris}, while additional {\em direct} couplings to light quarks are easily handled within our framework (see the Supplemental Material to this Letter).

We begin 
by noting that ALP-lepton couplings arise at the 3-loop order in this scenario, and therefore, are neglected throughout.
ALP couplings to quarks are generated by the ALP-gluon interactions.
Similarly, ALP-photon interactions are also generated by ALP-gluon interactions, though these are suppressed by $\mathcal{O}(\alphaem^2)$.

For low masses, ALP-gluon interactions can be studied using chiral perturbation theory~(\chipt), while for $m_a \gg \Lambda_{\rm QCD}$ perturbative QCD (pQCD) can be employed.
However,  no reliable calculations are available for most QCD-scale masses.
Furthermore, pQCD only predicts the total hadronic decay rate.
It does not inform experimenters which decays to look for, or how to determine the sensitivity of any exclusive decays.

Since $a\to\pi\pi$ and $a\to\pi^0\gamma$ are forbidden by $CP$ and $C$, respectively, the dominant hadronic decays for low-mass ALPs will be $a \to 3\pi^0$ and $a\to \pi^+\pi^-\pi^0$, even though they violate isospin, along with $a \to \pi^+\pi^-\gamma$, which is suppressed by a factor of \alphaem~\cite{Dobrescu:2000jt}.
The decay rates are similar for both $3\pi$ modes and
to leading order (LO) in \chipt are~\cite{Bauer:2017ris}
\begin{align}
	\label{eq:a3pi}
  	\Gamma_{a \to 3\pi}
	 \approx
	\frac{\pi m_a m_{\pi}^4 c_g^2 \delta_I^2}{\Lambda^2 f_{\pi}^2}
	\mathcal{K}_{3\pi}\!\left(\frac{m_{\pi}^2}{m_a^2} \right)
	 \,\, {\rm for}\,\, m_a \lesssim 1\,{\rm GeV},
\end{align}
where
%
%
$\delta_I \equiv (m_d - m_u)/(m_d + m_u) \approx \sfrac{1}{3}$
is the isospin violation induced by $m_u \neq m_d$ 
and $\mathcal{K}_{3\pi}$ contains the final-state kinematic factors (see Supplemental Material).
In the pQCD regime, the total rate to hadrons is $\Gamma_{a \to gg}$, which at one-loop order is\,\cite{Spira:1995rr}
\begin{align}
	\label{eq:agg}
  	\Gamma_{a \to gg }
 	\approx
  	\frac{32 \pi \alpha^2_s c_g^2 m_a^3}{ \Lambda^2}\!\left[1 \!+\! \frac{83\alpha_s}{4\pi}\right]  \, {\rm for}\,\, m_a \!\gg \!\Lambda_{\rm QCD}.
\end{align}
For $m_a \approx 2$\,GeV, the one-loop correction is comparable in size to the leading-order result, making this the smallest mass where Eq.~\eqref{eq:agg} has $\mathcal{O}(1)$ validity.
Naively, it is tempting to interpolate the total hadronic rate from where $a\to 3\pi$ is the dominant hadronic decay to where the pQCD result is valid;
however, even though such an interpolation only covers a factor of 4 in $m_a$,
numerically
\begin{align}
	\frac{\Gamma_{a \to gg}(m_a = 2\,{\rm GeV})}{\Gamma_{a \to 3\pi}(m_a = 0.5\,{\rm GeV})}
	\approx
	\mathcal{O}(10^5) \, !
\end{align}
Clearly a deeper understanding of the hadronic interactions of QCD-scale ALPs is required ---which is our primary focus.

By performing a chiral transformation of the light-quark fields~\cite{Georgi:1986df,Bardeen:1986yb,Krauss:1986bq},
we replace the $agg$ vertex by ALP-quark axial-current couplings,
which we subsequently match to the chiral Lagrangian.
This leads to ALP-$\pi^0$ kinetic mixing and ALP-$\eta^{(\prime)}$ kinetic and mass mixing 
making it possible to assign the ALP a $U(3)$ representation at low masses.
We assign all ALPs up to $\approx 3$\,GeV the $U(3)$ representation\footnote{Close to 3\,GeV mixing with the $\eta_c$ charmonium state should be considered. We leave this for future studies.}
\begin{align}
	\label{eq:a}
	\frac{f_{\pi}}{f_a}{\boldsymbol a} = \frac{f_{\pi}}{f_a} \frac{\tilde{\alpha}_s(m_a)}{\sqrt{6}}{\rm diag}\{\mathcal{C}_u, \mathcal{C}_d, \mathcal{C}_s \},
\end{align}
where $\mathcal{C}_q$ are $m_a$-dependent dimensionless constants,
$f_a \equiv -\Lambda/32\pi^2 c_g$
%
%
is the ALP decay constant, and
\begin{align}
  	\tilde{\alpha}_s(m_a) \equiv
		\begin{cases}
			1 & \!\!\!\!\text{for } m_a \leq 1\,{\rm GeV} \\
			\alpha_s(m_a) & \!\!\!\!\text{for } m_a > 1\,{\rm GeV}
			\end{cases} \,
\end{align}
accounts for $\alpha_s$ running which weakens ALP-gluon interactions at higher masses.\footnote{To obtain smooth results, we take $\alpha_s(1\,{\rm GeV})=1$, then interpolate to the known value for $m_a > 1.5$\,GeV.}
{\em N.b.}, we factored out $f_{\pi}/f_a$  to make this dependence explicit, and
follow the normalization convention
\begin{align}
	\label{eq:U3norm}
	\langle \boldsymbol{PP} \rangle \equiv 2{\rm Tr}[\boldsymbol{PP}]
	= 1\, ,
\end{align}
for the pseudoscalar $U(3)$ generators ${\boldsymbol \pi^0}$, ${\boldsymbol \eta}$, and ${\boldsymbol \eta'}$. 

\begin{figure}[t]
\includegraphics[width=0.5\textwidth]{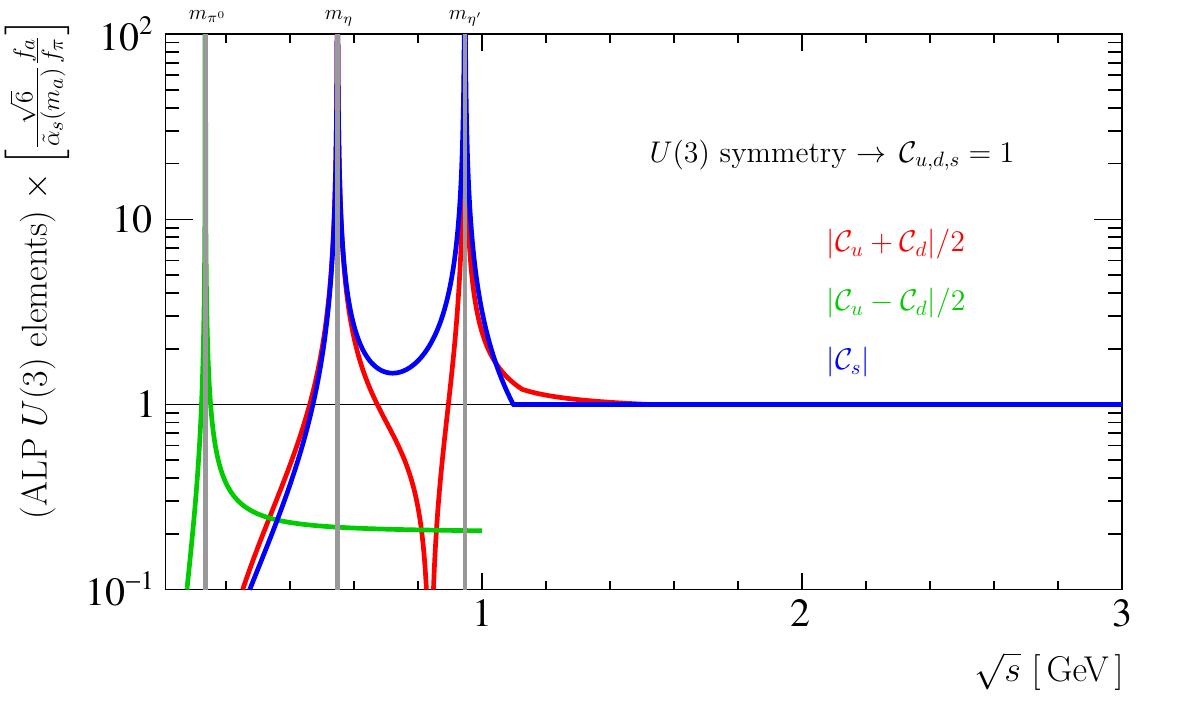}
\caption{ALP $U(3)$ representation. Since isospin-violating decays are small above $m_{\eta'}$, where the isospin-violating component is highly uncertain, we ignore such decays for $m_a > m_{\eta'}$.}
\label{fig:su3}
\end{figure}

For $m_a \lesssim 1$\,GeV, we derive the ALP-$P$ mixings, for $P = \pi^0, \eta, \eta'$, using the LO chiral Lagrangian by extending previous works, {\em e.g.}\ Ref.~\cite{Bauer:2017ris}, to three flavors and to higher order in $\delta_I$.
The full calculations are in the Supplemental Material.
Here, we provide simplified expressions to LO in $\delta_I$ and taking $m_s \gg m_d \approx 2 m_u$.
The ALP-$P$ kinetic and mass mixing cause the $P$ fields to pick up small admixtures of the physical ALP state and {\em vice versa}:
\begin{align}
	P &\approx	P_{\rm phy} + \frac{f_{\pi}}{f_a} \mixa{P} \, a_{\rm phy}\,, \\
	a &\approx a_{\rm phy} - \frac{f_{\pi}}{f_a}\sum_P \mixa{P} P_{\rm phy}\,. \nonumber
\end{align}
Therefore, the ALP $U(3)$ matrix is
\begin{align}
{\boldsymbol a} =  \mixa{\pi^0} {\boldsymbol \pi^0} \!+\!  \mixa{\eta} {\boldsymbol \eta} \! +\! \mixa{\eta'} {\boldsymbol \eta'}\,\,\,{\rm for}\,\,m_a \lesssim 1\,{\rm GeV},
\end{align}
where the ALP-$P$ mixing factors are
\begin{align}
	\label{eq:chiptmix}
	\mixa{\pi^0}  \approx \mathcal{N}_{\pi^0}\frac{\delta_I m_a^2}{m_a^2 - m_{\pi}^2}&, \,\,
	\mixa{\eta}  \approx  \mathcal{N}_{\eta} \left[ \frac{m_a^2 - m_{\pi^0}^2 /2}{m_a^2 - m_{\eta}^2} \right], \nonumber \\
	\mixa{\eta'}  \approx&  \mathcal{N}_{\eta'} \left[ \frac{m_a^2 - 2 m_{\pi^0}^2}{m_a^2 - m_{\eta'}^2} \right] \, ,
\end{align}
and $\mathcal{N}_{\pi^0,\eta,\eta'}=\frac{1}{2},\frac{1}{\sqrt{6}},\frac{1}{2\sqrt{3}}$ are the $\boldsymbol{P}$ normalization factors.
%
At high masses, the $U(3)$ symmetry is expected to be restored; thus the ALP $U(3)$ representation should be
\begin{align}
	\label{eq:Cuniv}
	\mathcal{C}_u \approx \mathcal{C}_d \approx \mathcal{C}_s \approx 1
	\, {\rm for}\,\,
	m_a \!\gg \!\Lambda_{\rm QCD}.
\end{align}
The $\mathcal{C}_q$ values obtained from Eq.~\eqref{eq:chiptmix} are close to unity near 1\,GeV; therefore, we interpolate between the low-mass and high-mass regions by setting each $\mathcal{C}_q$ element to unity once it intersects unity above $m_{\eta'}$ (see Fig.~\ref{fig:su3}).

When $m_a$ is in the non-perturbative regime of QCD, this $U(3)$-based representation is the most natural one, and
can be used to calculate the production and decay rates of ALPs.
Before moving onto such calculations, we stress that for $0.5 \lesssim m_a \lesssim 2$\,GeV there are $\mathcal{O}(1)$ uncertainties on $\boldsymbol{a}$.
Many LO \chipt predictions require $\mathcal{O}(1)$ corrections even for $\eta$ decays (see, {\em e.g.}, Ref.~\cite{Bijnens:2007pr}).
Furthermore, while parton-hadron duality is roughly valid above 1\,GeV for vector currents\,\cite{Ilten:2018crw}, not enough is known about $\eta^*$ states to assert that this holds to better than $\mathcal{O}(1)$ for ALPs. 
While the precision of $\boldsymbol{a}$ could be improved, 
adding direct quark couplings to the ALP model also induces $\mathcal{O}(1)$ changes in $\boldsymbol{a}$.
Therefore, a more natural approach is to adopt $\mathcal{C}_u$, $\mathcal{C}_d$, and $\mathcal{C}_s$ as effective ALP parameters, with the goal of experimentally exploring all $\mathcal{O}(1)$ deviations from the pure ALP-gluon model.

The interactions of pseudoscalar mesons are well described at low energies by
the hidden local symmetries framework of vector meson dominance~(VMD)\,\cite{Fujiwara:1984mp,Sakurai:1960ju}.
Due to ALP-pseudoscalar mixing, which generates the ALP $U(3)$ representation, we can also employ VMD to study ALP interactions.
However, since VMD only includes ground-state mesons, the effective theory breaks down once $m_a \gtrsim m_{\eta^*} \approx 1.5$\,GeV.
Ref.~\cite{Ilten:2018crw} showed how $e^+e^- \to V^{(*)}$ data can be used to predict the hadronic decay rates of any vector particle.
While no high-purity source of $P^{(*)}$ currents exists, with minimal assumptions we can also use $e^+e^-$ data to extend VMD-based pseudoscalar predictions up to 3\,GeV.

We begin by considering an interaction vertex with two vectors and one pseudoscalar ($VVP$).
The amplitude for the process $V_1(p_1) \!\to\! V_2(p_2) P(q)$ must be of the form
\begin{align}
  	\label{eq:amp_vvp}
	\mathcal{A}_{V_1\! \to \!V_2 P}
&
 =  	\epsilon_{\mu\nu\alpha\beta} \epsilon^{\mu}_1 \epsilon^{*\nu}_2 p_1^{\alpha} p_2^{\beta} \, \cF \! \left(p_1^2,p_2^2,q^2\right) \nonumber \\
& 	\qquad \qquad \times
	\frac{3 g^2}{4 \pi^2 f_{\pi}}\langle \boldsymbol{ V_1 V_2 P} \rangle \, ,
\end{align}
since this is the only valid Lorentz structure. The unknown function $\cF$ should satisfy
\begin{align}
	\cF \! \left(p_1^2,p_2^2,q^2\right)
\!=\!	\begin{cases}
  		\approx 1 & \!\!\!\text{for } m_1 \ll m_{V_1^*} \,\,\,\,\, (\rm VMD) \\
  		\propto \frac{1}{m_1^{4}} & \!\!\!\text{for } m_1 \gg \Lambda_{\rm QCD}\,\,({\rm pQCD})
	\end{cases} \, ,
\end{align}
where $m_1^2 = p_1^2$ and $m_{V_1^*}$ denotes the pole mass of the first excited vector meson with the same $U(3)$ representation as $V_1$.
The pQCD power-counting rule is $\mathcal{A} \propto m_1^{4-n}$, where $n$ is the number of partons involved in the vertex (6 for $VVP$)~\cite{Lepage:1980fj}.
Since for $m_1 \lesssim m_{V_1^*}$ $\mathcal{F}$ is approximately independent of the ground-state meson masses, we make the {\em ansatz}
\begin{align}
  	\cF \! \left(p_1^2,p_2^2,q^2\right) \to \cF(m_1) \, ,
\end{align}
which relies on $\cF$ being controlled by the heaviest dynamical scale, $m_1$ here, when all other masses are for ground-state mesons.
As shown in Ref.~\cite{Ilten:2018crw}, treating $e^+e^- \to q\bar{q}$ production as the sum of currents with $\rho$-like, $\omega$-like, and $\phi$-like $U(3)$ quantum numbers, rather than the sum of many $V^*$ resonances, provides a good description of the data for $m \equiv \sqrt{s} \gg m_{V^*}$.
Therefore, the $\mathcal{F}$ function can be extracted from data using
\begin{align}
  \mathcal{F}(m) \!\approx\!\! \left[ \frac{3 m \left[\frac{\sigma_{e^+e^- \to f}(m)}{\sigma_{e^+e^-\to\mu^+\mu^-}(m)}  \right]}{\Gamma_{V \to f}^{\rm VMD}(m)} \right]^{\frac{1}{2}} \!\!\!\!\!\!\times\!\!
\begin{cases}
  \sqrt{\frac{2}{3}}& \!\!\!(\rho{\rm -like}) \\
  \sqrt{6}& \!\!\!(\omega{\rm -like}) \\
  \sqrt{3}& \!\!\!(\phi{\rm -like})
\end{cases}
\end{align}
where $\Gamma_{V \to f}^{\rm VMD}(m)$ is the width obtained using VMD with $\mathcal{F} = 1$.

\begin{figure}[]
\includegraphics[width=0.5\textwidth]{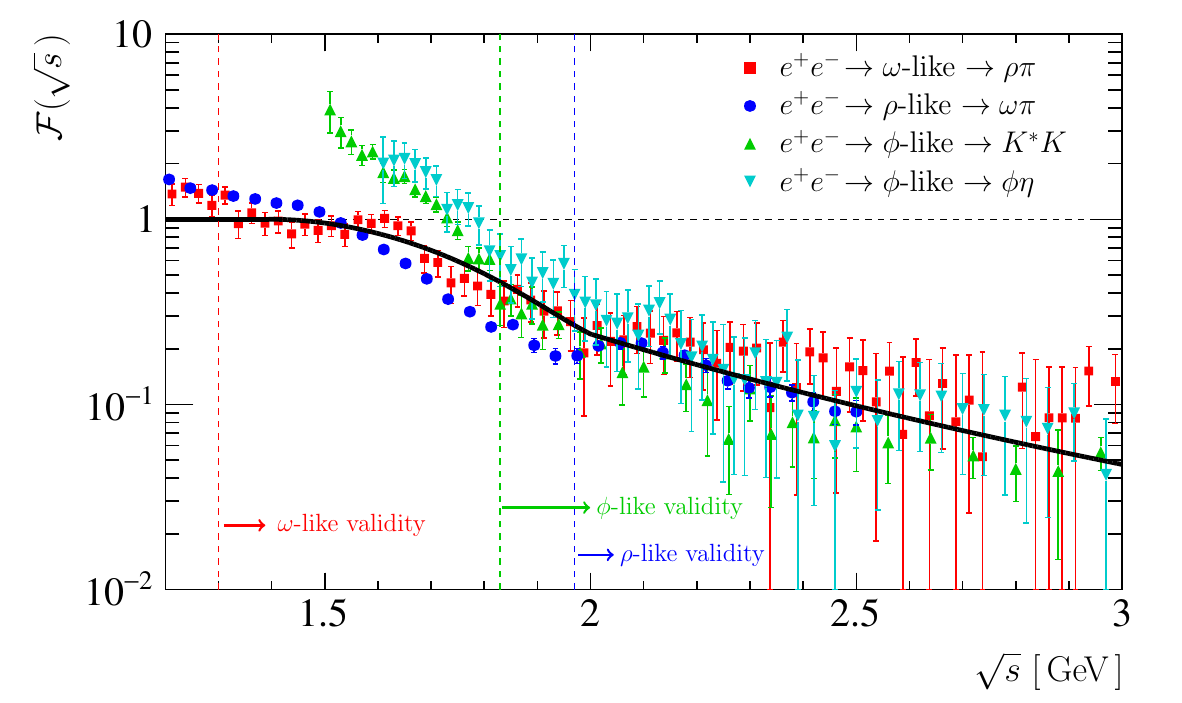}
\caption{$\mathcal{F}$ from Eq.~\eqref{eq:amp_vvp} determined from $e^+ e^-$ data\,\cite{TheBaBar:2017vzo,Aubert:2007ym,Aubert:2004kj}. Since we ignore resonance contributions, each result is only valid at masses where narrow resonance contributions are small. We define these as: ($\omega$-like) above where the sizable $\omega$--$\phi$ interference effect in the $3\pi$ final state becomes negligible, ($\rho$-like)~${m_a \gtrsim m_{\rho^*}+\Gamma_{\rho^*}}$, and ($\phi$-like) $m_a \gtrsim m_{\phi^*}+\Gamma_{\phi^*}$. }
\label{fig:eedata}
\end{figure}

Figure~\ref{fig:eedata} shows that all available $e^+e^- \to V_1 \to V_2P$ data are consistent with 
\begin{align}
  \label{eq:F}
\mathcal{F}(m) \!=\!\!
\begin{cases}
  1 & \text{for } m < 1.4\,{\rm GeV} \\
  \text{interpolation} & \text{for } 1.4 \leq\! m \!\leq 2\,{\rm GeV}\\
\left[\frac{\beta_{\mathcal{F}}}{m}\right]^4 & \text{for } m > 2\,{\rm GeV}
\end{cases}
\end{align}
where $\beta_{\mathcal{F}} = 1.4$\,GeV is determined from the data.
Furthermore, in the Supplemental Material we show that all $e^+e^- \to V \to PP$ data\,\cite{Lees:2012cj,Lees:2013gzt} are also consistent with Eq.~\eqref{eq:F}, modulo the pQCD power-law scaling is $m^{-3}$ due to the dimensionality of the VMD-based $VPP$ vertex.
Since $\mathcal{F}$ is simply a smooth monotonic transition from VMD to pQCD, we expect this function to be approximately valid for any 3-meson vertex where only the decaying particle is not a ground-state meson (corrected for vertex dimensionality if needed).
We will show below how to use Eq.~\eqref{eq:F} to extend VMD-based calculations up to 3\,GeV, and validate our approach using known $\eta_c$ and $\eta^*$ decay branching fractions.

The amplitude for $P \to V_1 V_2$ must have the same Lorentz structure as Eq.~\eqref{eq:amp_vvp},
and by crossing symmetry must share the same $\mathcal{F}$.
Therefore, using the standard VMD framework---but inserting $\mathcal{F}(m_P)$---we can calculate $\Gamma_{a \to VV}(m_a)$ up to $\approx 3$\,GeV.
These straightforward calculations follow directly from the standard VMD ones and are provided in the Supplemental Material.
Moreover,
using the same framework  we calculate $\Gamma_{\eta_c \to VV}$. 
Table~\ref{tab:VV} shows that our $\eta_c \to VV$ predictions are consistent with the experimental values to $\mathcal{O}(10\%)$.
Alternatively, $\Gamma_{\eta_c \to VV}$ can be calculated using pQCD; however, this approach underestimates the measurements\,\cite{Sun:2010qx} by $\mathcal{O}(10)$ even when including higher-twist effects (known as the $\eta_c \to VV$ puzzle).
That our predictions for $\Gamma_{\eta_c \to VV}$ achieve $\mathcal{O}(10\%)$ accuracy provides strong validation of the approach developed here.

\begin{table}[]
\centering

\begin{tabular}{c|ccc}
  {} & This Work & \multicolumn{2}{c}{Experiment} \\
{} & VMD$\times|\mathcal{F}(m)|^2$ & PDG & $SU(3)$  \\
\hline
$\mathcal{B}(\eta_c \to \rho\rho)$ & 1.0\% & $1.8 \pm 0.5\%$ & $1.10\pm0.14\%$ \\
$\mathcal{B}(\eta_c \to \omega\omega)$ & 0.40\% & $0.20 \pm 0.10\%$ & $0.44\pm0.06\%$\\
$\mathcal{B}(\eta_c \to \phi\phi)$ & 0.25\% &  $0.28 \pm 0.04\%$ & $0.28\pm0.04\%$\\
$\mathcal{B}(\eta_c \to K^{*}\overline{K}{}^*)$ & 0.91\% &  $0.91 \pm 0.26\%$ & $1.00\pm0.13\%$\\
\hline
\end{tabular}
\caption{Validation using $\eta_c \to VV$ decays:
	Our predictions are consistent with the PDG average of each experimental value\,\cite{PDG,Liu:2012eb}.
	Furthermore, we derive more precise experimental values by averaging the PDG $\eta_c \to VV$ results assuming $SU(3)$ symmetry in these decays (the $SU(3)$ column), and find that our predictions are consistent with these $SU(3)$-averaged experimental results to $\mathcal{O}(10\%)$.
}
\label{tab:VV}
\end{table}

\begin{figure}[t]
  \includegraphics[width=0.49\textwidth]{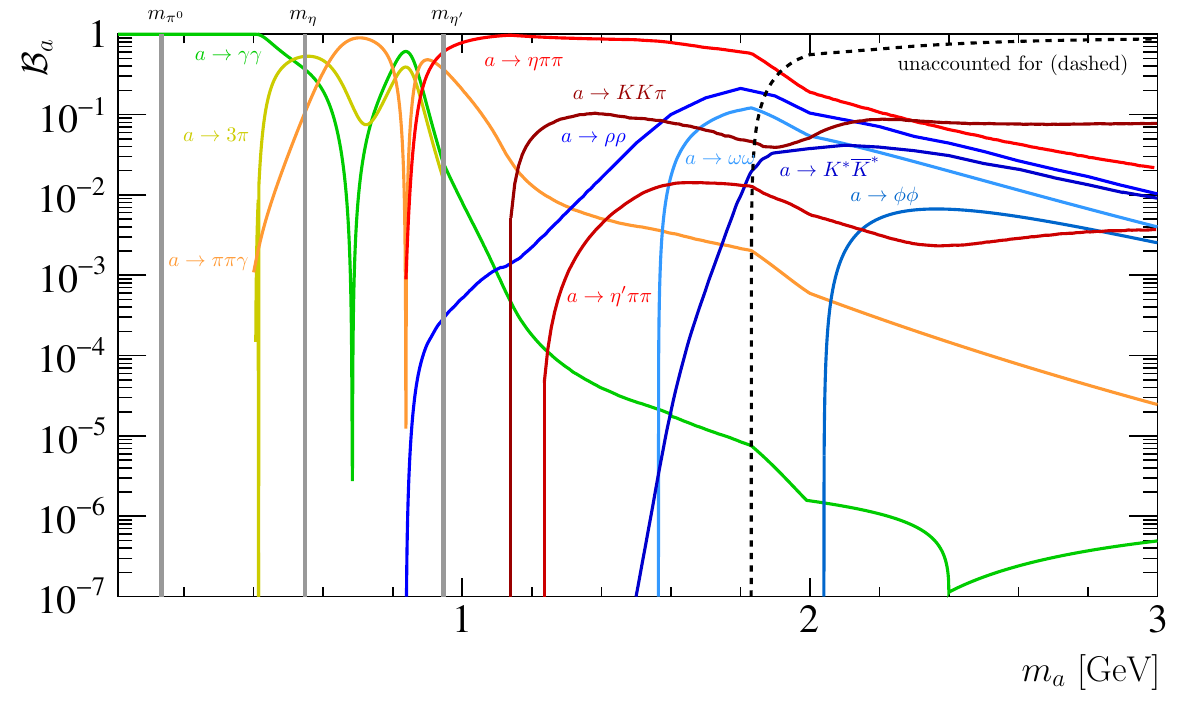}
\caption{ALP decay branching fractions to all final states considered; decay widths are given in the Supplemental Material.}
\label{fig:decays}
\end{figure}

Given any ALP $U(3)$ representation and the mass-dependent vertex scaling function Eq.~\eqref{eq:F}, we can calculate exclusive hadronic ALP decay widths and its total hadronic width.
Here we summarize our calculations for the representation shown in Fig.~\ref{fig:su3}, while the details are provided in the Supplemental Material. \\ \vspace{-1.5em}
\begin{itemize}[leftmargin=0.0em,itemindent=1.0em,itemsep=0.0em]
  \item $\Gamma_{a \to VV}$: As discussed above, we calculate $a\to\rho\rho$, $a \to \omega\omega$, $a\to\phi\phi$, and $a\to K^* \overline{K}{}^*$ using our extended-VMD framework.
	Schematically, the $a\to VV$ and VMD-based $\eta' \to VV$ rates are related via
	\begin{align}
		\label{eq:aVVrate}
		\Gamma_{a \to VV} = \left|  \frac{f_{\pi} \mixa{\{V,V\}} \mathcal{F}(m_a) }{f_a \langle \boldsymbol{\eta^{\prime}\{V,V\}} \rangle} \right|^2 \Gamma_{\eta^{\prime} \to VV}^{m_{\eta'} \to m_a}\,.
	\end{align}
	Additionally, we calculate $\Gamma_{a \to \pi\pi\gamma}$ as $a \to \rho\rho$ followed by $\rho$--$\gamma$ mixing and $\rho\to\pi\pi$.
  \item $\Gamma_{a \to VP}$: Since $a \to \rho\pi$ violates isospin and $a \to K^*K$ violates $SU(3)$ symmetry, these are subleading and difficult to calculate; thus we do not consider them.\footnote{Determining the $U(3)$-violating components of $\boldsymbol{a}$ and the kaon-loop contributions to isospin-violating final states would be tedious. None of these $U(3)$-violating modes are expected to be important at any masses.} Most other $a \to VP$ decays involving ground-state mesons violate $C$, so also are not considered.
  \item $\Gamma_{a \to \gamma\gamma}$: The $a \to \gamma\gamma$ decay rate is given by
	\begin{align}
		\Gamma_{a \to \gamma\gamma} = \frac{\alpha_{\rm EM}^2 m_a^3}{(4\pi)^3 f_a^2} \left| \mathcal{C}^{\chi}_{\gamma} +  \mathcal{C}_{\gamma}^{\rm VMD} + \mathcal{C}_{\gamma}^{{\rm pQCD}} \right|^2,
	\end{align}
	where at low masses $\mathcal{C}^{\chi}_{\gamma} \approx 1$ is generated by the chiral transformation, while at high masses pQCD quark-loop contributions (at two-loop order) are important~\cite{Bauer:2017ris}.
	Calculated for the first time here from ${a \to VV \to \gamma\gamma}$ with $V$--$\gamma$ mixing,
	\begin{align}
 \mathcal{C}_{\gamma}^{\rm VMD} \!\!&=  \!\!-\mathcal{F}(m_a) \!\! \left[ 3 \mixatwo{\rho}{\rho}
\!+\! \frac{1}{3} \mixatwo{\omega}{\omega}
\!+\! \frac{2}{3} \mixatwo{\phi}{\phi}
\!+\! 2  \mixatwo{\rho}{\omega} \right] \nonumber \\
 &= -\mathcal{F}(m_a) \frac{2 \tilde{\alpha}_s(m_a)}{3\sqrt{6}}\left(4\,\mathcal{C}_u + \mathcal{C}_d + \mathcal{C}_s  \right),
	\end{align}
	is found to be the dominant contribution over most of the mass range considered.
	{\em N.b.}, each contribution is turned on/off for $m_a$ values where it is either invalid or where double counting of contributions would occur.

  \item $\Gamma_{a \to 3\pi}$:  We calculate these rates using the LO chiral Lagrangian, and add a data-derived $k$-factor to account for final-state-pion rescattering effects.
	We only consider these decays up to $m_{\eta'}$, since at higher masses this $k$-factor is no longer reliable.
	We consider isospin-violating $a$--$\pi^0$ mixing, and our calculation is the first to consider $a$--$\eta^{(\prime)}$ mixing followed by ${\eta^{(\prime)}\to 3\pi}$.
	We leave a detailed presentation to the Supplmental Material.
  \item $\Gamma_{a \to PPP}$: The amplitudes for $a\to \eta^{(\prime)}\pi\pi$ and ${a \to K\overline{K}\pi}$ are dominated by scalar and tensor resonances.
	Specifically, for  $a\to \eta^{(\prime)}\pi\pi$ we consider ${a \to \sigma(\pi\pi)\eta^{(\prime)}}$, $a \to f_0(\pi\pi)\eta^{(\prime)}$, $a \to a_0(\eta^{(\prime)}\pi)\pi$, ${a \to f_2(\pi\pi)\eta^{(\prime)}}$, and a contact term.
	For $a \to K\overline{K}\pi$ we consider $a \to S_{K\pi}(K\pi)K$, where the $K\pi$ $S$-wave amplitude is taken from Ref.~\cite{Lees:2015zzr}, and $a \to a_0(KK)\pi$.
	Schematically, the $a \to PPP$ and $\eta' \to PPP$ amplitudes are related similarly to Eq.~\eqref{eq:aVVrate}, {\em e.g.},
	\begin{align}
		\mathcal{A}_{a \to f_0(\pi\pi)\eta} = \frac{f_{\pi} \mixa{\eta f_0} \mathcal{F}(m_a)}{f_a \langle \boldsymbol{\eta^{\prime} \eta f_0} \rangle } \mathcal{A}_{\eta' \to f_0(\pi\pi)\eta}^{m_{\eta'} \to m_a} \,\,.
	\end{align}
	All scalar resonance amplitudes are taken from the ${\eta' \to \eta\pi\pi}$ model of Ref.~\cite{Fariborz:1999gr}, where they were determined by fitting all available data.
	We use a similar approach to derive the $f_2(1270)$ tensor-meson contribution in the Supplemental Material.
	Unlike above, we cannot obtain the $\mathcal{F}$ functions for these vertices directly from data.
	Given that the dimensionality of each of these vertices is the same as that of $VVP$, we also use Eq.~\eqref{eq:F} here.
	This {\em universality} assumption is validated by the fact that we accurately predict both $\mathcal{B}(\eta_c \to \eta\pi\pi)$ and $\mathcal{B}(\eta(1760)\to\gamma\gamma)\times\mathcal{B}(\eta(1760)\to\eta'\pi\pi)$ to $\approx 20\%$,
	and $\mathcal{B}(\eta_c \to K\overline{K}\pi)$ to $\approx 10\%$.
	Given that $a \to \eta\pi\pi$ or $a\to K\overline{K}\pi$ has the largest branching fraction for $m_a \gtrsim 1$\,GeV, the lack of more stringent data-driven constraints here is the weakest component of our calculations, though these data-driven tests suggest that the uncertainties are small.
	(These predictions could be improved with a better experimental understanding of the excited $\eta^*$ states.)  
  \item $\Gamma_{a\to gg}$: The NLO pQCD calculation of Eq.~\eqref{eq:agg} derived in Ref.~\cite{Bauer:2017ris} is adopted here.
	\item $\Gamma_a$ (total hadronic width): We take $\Gamma_a = \Gamma_{a\to gg}$ for $m_a \gtrsim 1.84$\,GeV, while for lower masses, the sum of all exclusive modes is used for $\Gamma_a$. At $m_a \simeq 1.84\,\rm{GeV}$ we find $\Gamma_{a\to gg} \approx \underset{i=\rm exc.}{\sum} \Gamma_{i}$.
\end{itemize}
The decay branching fractions are summarized in Fig.~\ref{fig:decays}.
The unaccounted for branching fraction is also shown, and is substantial for $m_a \gtrsim 2$\,GeV.
This includes decays such as $a\to AA$, {\em i.e.}\ two axial-vector mesons, which should be comparable to $a\to VV$ above about $2.5$\,GeV, and many decay paths that involve excited resonances, rescatterings, {\em etc.}
For example $\mathcal{B}(\eta_c \to 6\pi) \approx 20\%$ so we expect ALP decays to many-body final states to be at about the same rate.
We stress that unaccounted for decay modes should only be important for ALP masses where $\Gamma_a \approx \Gamma_{a\to gg}$; therefore, our predictions for the total hadronic width---and the ALP lifetime---should not be affected by unaccounted for decays.

When evaluating the constraints on this model,
we focus on the $m_{\pi} < m_a < 3$\,GeV region, where our work has the biggest impact.
Constraints where $f_a \lesssim 3 f_{\pi}$ are omitted, {\em e.g.}, bounds from radiative $J/\psi$ decays, since we assumed $f_{\pi} \ll f_a$ when deriving $\boldsymbol{a}$.
Details on all calculations are provided in the Supplemental Material, while in Fig.~\ref{fig:limits} and below we summarize the constraints.
\begin{itemize}[leftmargin=0.0em,itemindent=1.0em,itemsep=0.0em]
\item We recast existing limits on the $a\gamma\gamma$ vertex from LEP\,\cite{Abbiendi:2002je,Knapen:2016moh} and beam-dump experiments\,\cite{Bjorken:1988as,Blumlein:1990ay,Ariga:2018uku} using our $\mathcal{B}(a \to \gamma\gamma)$ result and our $a\to\gamma\gamma$ calculation to relate the $a\gamma\gamma$ interaction strength to $f_a$. In Ref.~\cite{Aloni:2019ruo}, we derive new constraints using $\gamma p \to p a(\gamma\gamma)$ data from {\sc GlueX}~\cite{AlGhoul:2017nbp}.
\item We derive new constraints from $\phi \to \gamma a(\pi\pi\gamma,\eta\pi^0\pi^0)$ and  $\eta' \to \pi^+\pi^-a(\pi^+\pi^-\pi^0)$.
We are not aware of any bump hunts here, and instead assume that the entire known branching fractions to these final states~\cite{PDG} are due to ALPs. Clearly dedicated searches would be much more sensitive.
\item We derive new constraints from  $b\to sa$ penguin decays.
At one loop, the $agg$ vertex generates an axial-vector $a tt$ coupling~\cite{Bauer:2017ris} resulting in enhanced rates for $B \to K^{(*)}a$ decays~\cite{Batell:2009jf,Hiller:2004ii,Bobeth:2001sq,Choi:2017gpf}.
The loop contains a UV-dependent factor\,\cite{Freytsis:2009ct} schematically given by ${\approx [\log{\Lambda^2_{\rm UV}/m^2_t} \pm \mathcal{O}(1)]}$, which we take to be unity (corresponding to an $\mathcal{O}({\rm TeV})$ UV scale).
This induces $\mathcal{O}(1)$ arbitrariness on the following constraints:
\begin{itemize}[leftmargin=0.5em]
\item The published $m_{\eta\pi\pi}$ spectrum of Ref.~\cite{Aubert:2008bk} is used to constrain $\mathcal{B}(B^{\pm} \to K^{\pm} a) \times \mathcal{B}(a \to \eta \pi^+\pi^-)$ for ${m_a < 1.5}$\,GeV, excluding the $\eta'$ peak region.
\item The published $m_{K^*K}$ spectrum of Ref.~\cite{Aubert:2008bk} is used to constrain $\mathcal{B}(B^{\pm} \to K^{\pm} a) \times \mathcal{B}(a \to K^{\pm}K_S\pi^{\mp})$ for ${0.85<m_{K\pi}<0.95\,{\rm GeV}}$ and $m_a < 1.8$\,GeV.
\item The known value of $\mathcal{B}(B^0 \to K^0 \phi\phi)$\,\cite{Lees:2011zh} is used to constrain $\mathcal{B}(B^0 \to K^0 a) \times \mathcal{B}(a \to \phi \phi)$ assuming the entire decay rate is due to ALPs.
\item The known value of $\mathcal{B}(B^{\pm} \to K^{\pm} \omega(3\pi))$ is used to constrain $\mathcal{B}(B^{\pm} \to K^{\pm} a) \times \mathcal{B}(a \to \pi^+\pi^-\pi^0)$ for ${0.73 < m_a < 0.83}$\,GeV, which is the $3\pi$ mass window shown in Ref.~\cite{Chobanova:2013ddr}, assuming the entire decay rate is due to ALPs.
\item Since the ALPs considered here are not massive enough to decay into charm hadrons, the observed inclusive $b \to c$ branching fraction\,\cite{PDG} is used to place an upper limit on the inclusive $b \to s a$ rate of ${\mathcal{B}(b \to s a)< \left[1 - \mathcal{B}(b \to c)\right]}$.
\end{itemize}
\item Similarly, we recast existing limits on ALP--$W/Z$ couplings from Ref.~\cite{Izaguirre:2016dfi} using the $s\to d$ penguin decays $K^{\pm} \to \pi^{\pm}\gamma\gamma$\,\cite{Ceccucci:2014oza} and $K_L \to \pi^{0}\gamma\gamma$\,\cite{Abouzaid:2008xm} and the same UV-completion assumptions.
\end{itemize}
Over much of the considered mass range the constraints on $\Lambda$ are below a TeV.
We stress that many of these constraints would be much stronger if dedicated searches were performed, {\em e.g.}, searches for $B \to K^{(*)}a$ with $a \to \gamma\gamma,\, 3\pi,\, \eta\pi\pi,\, K\overline{K}\pi,\, \rho\rho$, {\em etc.}\ would be incredibly powerful probes of QCD-scale ALPs---and could be performed with data already collected by LHCb.

\begin{figure}[!t]
\includegraphics[width=0.49\textwidth]{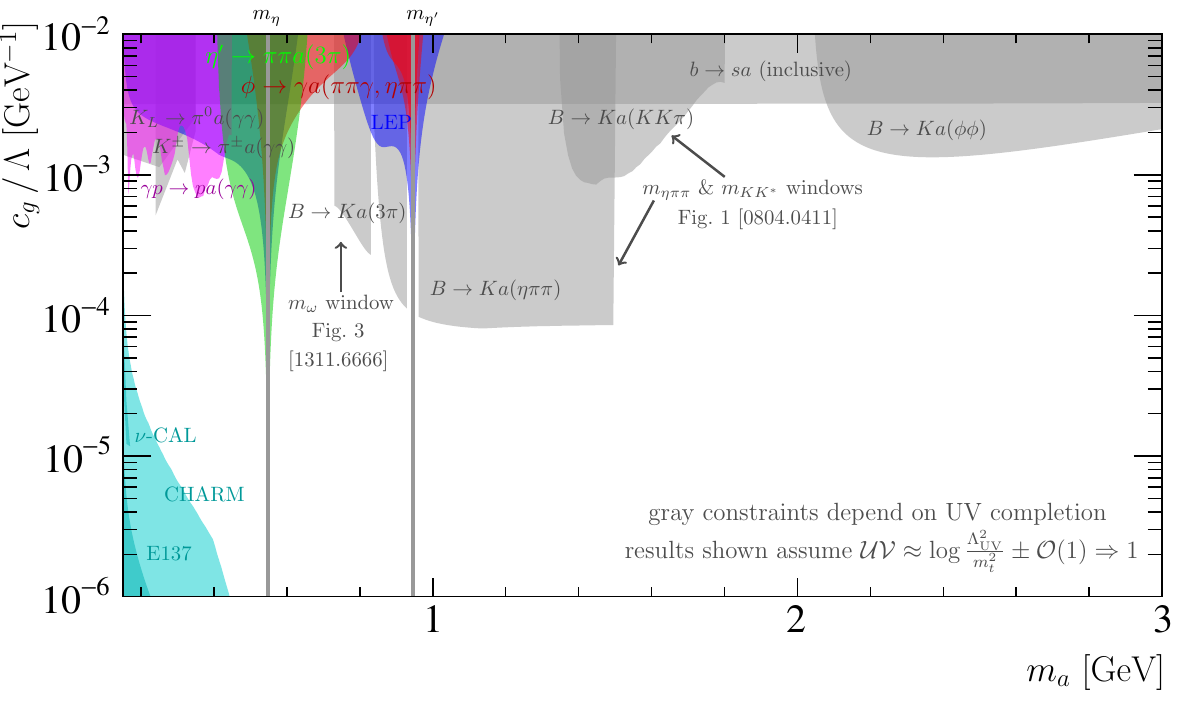}
\caption{Constraints on the ALP-gluon coupling.}
\label{fig:limits}
\end{figure}

In summary,
we presented a novel data-driven method for determining the hadronic interaction strengths of ALPs with QCD-scale masses.
Our method makes it possible to calculate the hadronic production and decay rates of ALPs, along with many of the largest ALP decay branching fractions to exclusive final states.
To illustrate the impact on QCD-scale ALP phenomenology, we considered the scenario where the ALP-gluon coupling is dominant over the ALP coupling to photons, electroweak bosons, and all fermions, but emphasized that our method is easily generalized to any set of ALP couplings to SM particles.
We showed that the constraints on this type of ALP are weak, though we also highlighted some promising searches that could provide improved sensitivity to QCD-scale ALPs, {\em e.g.}\ at LHCb.
Finally, our work determined the relationship between the ALP lifetime and its gluonic coupling, which is vital for studying the sensitivity of long-lived particle experiments~\cite{,Beacham:2019nyx}.

\begin{acknowledgments}
We thank Jesse Thaler for providing feedback on the early stages of this work and for providing comments on this Letter, along with Will Detmold,  Michele Papucci, and Maxim Pospelov for providing feedback on this Letter, and David Curtin, Phil Ilten, and Jure Zupan for useful discussions.
YS and MW performed part of this work at the Aspen Center for Physics, which is supported by U.S.\ National Science Foundation grant PHY-1607611. This work was supported by: YS was supported by the Office of High Energy Physics of the U.S.\ Department of Energy under grant contract number DE-SC00015476; and MW was supported by the U.S.\ National Science Foundation under contract number PHY-1607225.
\end{acknowledgments}

\twocolumngrid
\vspace{-8pt}
\section*{References}
\vspace{-10pt}
\def\bibsection{}
\bibliographystyle{utphys}
\bibliography{agg_bib}


\clearpage
\newpage
\maketitle
\onecolumngrid

\begin{center}
\textbf{\large Coupling QCD-scale axion-like particles to gluons} \\
\vspace{0.05in}
{ \it \large Supplemental Material}\\
\vspace{0.05in}
{Daniel Aloni, Yotam Soreq, and Mike Williams}
\end{center}

\onecolumngrid
\setcounter{equation}{0}
\setcounter{figure}{0}
\setcounter{table}{0}
\setcounter{section}{0}
\setcounter{page}{1}
\makeatletter
\renewcommand{\theequation}{S\arabic{equation}}
\renewcommand{\thefigure}{S\arabic{figure}}
\renewcommand{\thetable}{S\arabic{table}}
\newcommand\ptwiddle[1]{\mathord{\mathop{#1}\limits^{\scriptscriptstyle(\sim)}}}

\section{Details on ALP Theory}

\subsection{ALP effective Lagrangian}

Here we give more details about the axion-like particle model that is described in the Letter.
We start with the following effective Lagrangian:
\begin{align}
	\label{eq:LeffALP}
	\cL_{\rm eff}
=&	\bar{q}\left( i\slashed{D} -m_q \right)q
	+\frac{1}{2}\partial^\mu a \partial_\mu a
	-\frac{1}{2} m^2_a a^2
	+ \frac{c_{\gamma} 4\pi \alpha_{\rm EM}}{\Lambda} a F^{\mu\nu}\tilde{F}_{\mu\nu}
	- \frac{c_{g} 4\pi \alpha_{s}}{\Lambda} a G^{\mu\nu}\tilde{G}_{\mu\nu}
	- \frac{c_{q} }{2\Lambda} (\partial^\mu a) \bar{q}\gamma_\mu \gamma_5 q \, ,
\end{align}
where $\Lambda = -32\pi^2 c_g f_a$.
For $m_a \lesssim 4\pi f_\pi$, the Lagrangian of Eq.~\eqref{eq:LeffALP} can be matched to the Chiral Lagrangian~\cite{Georgi:1986df,Bardeen:1986yb,Krauss:1986bq}, see Eq.~\eqref{eq:LeffChiral} below.
In the Letter, we focus on the specific case where $c_{\gamma} = c_q = 0$; however, in this section, we keep these terms so that it is clear how to use our framework to obtain results for other ALP models.

Generically, both kinetic and mass mixing occur between the ALP and the neutral pseudoscalar mesons $P=\pi^0$, $\eta$, and $\eta'$.
Both the kinetic and the mass mixing terms scale as
\begin{align}
	\epsilon \equiv \frac{f_\pi}{f_a}\ll1 \, ,
\end{align}
which is the expansion parameter.
The mixing can be expressed as
\begin{align}
	\cL_{\rm mix}
=	\frac{1}{2} \partial_\mu\Phi_i K_{ij} \partial^\mu\Phi_j - \frac{1}{2}\Phi_i M^2_{ij} \Phi_j \, ,
\end{align}
where $\Phi = (a, \pi^0, \eta, \eta')$ and
\begin{align}\label{eq: K and M matrices}
	K
=	\begin{pmatrix}
	1 & - \epsilon K_{a\pi} & -\epsilon K_{a\eta} & -\epsilon K_{a\eta'} \\
	- \epsilon K_{a\pi} & 1 & 0 & 0 \\
	- \epsilon K_{a\eta} & 0 & 1  & 0 \\
	- \epsilon K_{a\eta'} & 0 & 0 & 1
	\end{pmatrix} \, ,  \qquad\qquad
	M^2
=	\begin{pmatrix}
	m_a^2 & \epsilon M_{a\pi}^2 & \epsilon M_{a\eta}^2 & \epsilon M_{a\eta'}^2 \\
	\epsilon M_{a\pi}^2 & M^2_{\pi\pi} & \delta_I M_{\pi \eta}^2 & \delta_I M_{\pi \eta^\prime}^2 \\
	\epsilon M_{a\eta}^2 & \delta_I M_{\pi \eta}^2 & M^2_{\eta\eta} & 0 \\
	\epsilon M_{a\eta'}^2 &\delta_I M_{\pi \eta^\prime}^2 & 0 & M^2_{\eta'\eta'}
	\end{pmatrix}
	\, .
\end{align}
To leading order in \chipt,  $M_{\pi \eta}^2/\sqrt{2} = M_{\pi \eta^\prime}^2=  -M_{\pi\pi}^2 / \sqrt{3}$ and diagonal elements of $M^2_{PP}$ are the physical masses. The model-dependent parameters $K_{aP}$ and $M^2_{aP}$ encode the information about the underlying ALP interactions with quarks and gluons.

To first order in $\epsilon$, the ALP and $P$ masses remain unchanged by the mixing, {\em i.e.}\ $m_P\approx M_{PP}$.
However, after a shift for the canonical kinetic term and the mass terms are diagonalized, the ALP field is redefined as
\begin{align}
	a
=& 	a_{\rm phy} - \epsilon \sum_P h(a,P,m_P) P_{\rm phy} \, ,
\end{align}
where the function $h$ is defined as
\begin{align}
	h(a,P,m_X)
	\equiv&
	\frac{1}{m_a^2 - m_P^2}\left[M^2_{aP} +  m^2_X  K_{aP}
	+ \delta_I \sum_{P^\prime}  M^2_{PP'}\frac{M_{a P^\prime}^2 + m_X^2 K_{a P^\prime}}{m_X^2 - m_{P^\prime}^2}	\right]
	\, .
\end{align}
This shift only affects ALP--ALP interactions, so we ignore it below.
The pseudoscalar fields become
\begin{align}
	\label{eq:Pphy}
	P
= 	P_{\rm phy} - \delta_I \sum_{P^\prime}  S_{PP'} P_{\rm phy}^\prime + \epsilon \,  \mixa{P} \, a_{\rm phy}
\end{align}
with
\begin{align}
	S_{PP'} = \frac{M_{P P^\prime}^2}{m_P^2 - m_{P^\prime}^2} \, , \qquad
	\mixa{P} = h(a,P,m_a) \, ,
\end{align}
which induces ALP--$P$ interactions.
Therefore, the ALP can be represented by the $U(3)$ matrix
\begin{align}
    \label{eq:aSU3}
{\boldsymbol a} =  \mixa{\pi^0} {\boldsymbol \pi^0} +  \mixa{\eta} {\boldsymbol \eta}  + \mixa{\eta'} {\boldsymbol \eta'},
\end{align}
where to leading order in isospin breaking (note that we consider this limit everywhere except for $a\to 3\pi$)
\begin{align}
  	\label{eq:aPSU3}
  	\mixa{P}
	\equiv
	2 {\rm Tr}[ \boldsymbol{a P}]
=	h(a,P,m_a)\Big|_{\delta_I\to0}
	\!\!\approx
	\frac{ M^2_{aP} +  m^2_a  K_{aP}}{m_a^2 - m_P^2}
\end{align}
and the $U(3)$ pseudoscalar meson generators are
\begin{align}
  {\boldsymbol \pi^0} = \frac{1}{2}{\rm diag}\{ 1,-1,0  \}\, , \
  {\boldsymbol \eta} = \frac{1}{\sqrt{6}}{\rm diag}\{ 1,1, -1  \}  \, ,\
    {\boldsymbol \eta'} = \frac{1}{2\sqrt{3}}{\rm diag}\{ 1,1, 2 \} \, ,
\end{align}
using $\sin{\theta_{\eta\eta'}} \approx -1/3$ and $\cos{\theta_{\eta\eta'}} \approx 2\sqrt{2}/3$.
%
We note that these mixing-angle values, which are inconsistent with more recent high-precision studies (though accurate enough for our purposes), were chosen as they lead to greatly simplified expressions in the following section.

\subsection{ALP couplings to hadrons and its low-mass $U(3)$ representation}

In this section, we determine the various mixing factors. Following Refs.~\cite{Bauer:2017ris,Alves:2017avw,Georgi:1986df}, we start with Eq.~\eqref{eq:LeffALP} and consider only $u,d,s$ quarks and define
\begin{align}
	{\boldsymbol m} = {\rm diag}\{ m_u,m_d,m_s \} \,\,{\rm and}\,\, {\boldsymbol Q} = \frac{1}{3}{\rm diag}\{2,-1,-1 \}\, .
\end{align}
We now preform the following chiral rotation to the quark fields, which ensures that the $agg$ vertex vanishes:
\begin{align}
	\label{eq:qrot}
	q
\ \to \ \exp\left[  i (a/f_a)  \kappa_q \gamma_5 \right] q
\end{align}
where $f_a$ is the ALP decay constant. 
In order to avoid mass mixing between the ALP and non-singlet $U(3)$ pseudoscalar states, namely ${\boldsymbol \pi^0}$ and ${\boldsymbol \eta_8}$, we choose
\begin{align}
	{\boldsymbol \kappa} = \frac{{\boldsymbol m}^{-1}}{\langle {\boldsymbol m}^{-1} \rangle}.
\end{align}
The rotation of Eq.~\eqref{eq:qrot} leads to
\begin{align}
	\label{eq:LeffALPRot}
	\cL_{{\rm eff},a}
=&	\bar{q}\left[ i\slashed{D} -\hat{m}_q(a) \right]q
	+\frac{1}{2}\partial^\mu a \partial_\mu a
	-\frac{1}{2} m^2_a a^2
	+ \frac{\hat{c}_{\gamma} }{4 \Lambda} a F^{\mu\nu}\tilde{F}_{\mu\nu}
	+ \frac{\left( \hat{c}_q + \kappa_q \right)}{f_a} (\partial^\mu a) \bar{q}\gamma_\mu \gamma_5 q \, ,
\end{align}
with
\begin{align}
	\label{eq:c_eff}
	\hat{m}_q(a)
=& 	\exp\left[  i (a/f_a)  \kappa_q  \gamma_5 \right]  m_q \exp\left[  i (a/f_a)  \kappa_q  \gamma_5 \right]
 \, ,  \nonumber \\
	\hat{c}_{\gamma}
=&	16\pi\alpha_{\rm EM} \left( c_{\gamma} - 2 N_c \langle \boldsymbol{\kappa QQ} \rangle c_{g} \right) \, , \\
	\hat{c}_q
=& 	\frac{c_{q}}{64\pi^2 c_g}\,, \nonumber
\end{align}
where $N_c = 3$ is the number of colors.

Next, following Ref.~\cite{Georgi:1986df}, we match Eq.~\eqref{eq:LeffALPRot} to the Chiral Lagrangian which gives
\begin{align}
	\label{eq:LeffChiral}
	\cL_{{\rm eff},a}
=&	\frac{f^2_\pi}{8} \langle  D^\mu \boldsymbol{\Sigma} D_\mu \boldsymbol{\Sigma}^\dagger \rangle
	+\frac{f^2_\pi}{4}B_0 \langle  \boldsymbol{\Sigma \hat{m}}^\dagger + \boldsymbol{\hat{m}\Sigma}^\dagger  \rangle
	-\frac{1}{2}m^2_0 \eta^2_0
	+ i \frac{f^2_\pi}{4 f_a} (\partial^\mu a)  \langle \left(\boldsymbol{\kappa}   +
	 \boldsymbol{\hat{c}} \right)  \left(\boldsymbol{\Sigma}^\dagger D_\mu \boldsymbol{\Sigma}  - \boldsymbol{\Sigma} D_\mu \boldsymbol{\Sigma}^\dagger\right) \rangle  \nonumber\\
&	+\frac{1}{2}\partial^\mu a \partial_\mu a -\frac{1}{2} m^2_a a^2
	+ \frac{\hat{c}_{\gamma}}{4\Lambda} a F^{\mu\nu}\tilde{F}_{\mu\nu}
	+ \cL_{\rm VMD}    \, ,
\end{align}
where $ B_0 = m^2_{\pi^0}/(m_u+m_d)  $, $f_\pi\approx93\,$MeV, $m^2_0$ is a hard breaking term due to the anomalous $U(1)$ symmetry which fixes the $\eta-\eta^\prime$ mixing angle $\theta_{\eta\eta^\prime}$, $\eta_0$ is the $U(1)$ Goldstone boson before this rotation, and we replace $\hat{m}_q(a)$ by its eigenvalue $\boldsymbol{\hat{m}}
= 	\exp\left[  i (a/f_a)  \kappa_q  \right]  m_q \exp\left[  i (a/f_a)  \kappa_q  \right] $.
We take the VMD term from Ref.~\cite{Fujiwara:1984mp} using
\begin{align}
	{\boldsymbol \Sigma}
= 	\exp\left( i 2 {\boldsymbol P} / f_\pi \right) \,  , \qquad\qquad
	D_\mu {\boldsymbol \Sigma}
= 	\partial_\mu {\boldsymbol \Sigma} + i e A_\mu [{\boldsymbol Q}, {\boldsymbol \Sigma} ]  \, ,
\end{align}
where the pseudoscalar and vector meson $U(3)$ matrices are
\begin{align}
	\label{eq:PSU3}
	{\boldsymbol P}
=	\frac{1}{\sqrt{2}}\begin{pmatrix}
	\frac{\pi^0}{\sqrt{2}} + \frac{\eta}{\sqrt{3}} +\frac{\eta'}{\sqrt{6}} & \pi^+ & K^+ \\
	\pi^- & -\frac{\pi^0}{\sqrt{2}} + \frac{\eta}{\sqrt{3}} +\frac{\eta'}{\sqrt{6}} & K^0 \\
	K^- & \bar{K}^0 & -\frac{\eta}{\sqrt{3}} + \frac{2\eta'}{\sqrt{6}}
	\end{pmatrix} \, , \qquad
	{\boldsymbol V}
=	\frac{1}{\sqrt{2}}\begin{pmatrix}
	\frac{\rho^0 + \omega}{\sqrt{2}} & \rho^+ & K^{*+} \\
	\rho^- & \frac{-\rho^0 + \omega}{\sqrt{2}} & K^{*0} \\
	K^{*-} & \bar{K}^{*0} & \phi
	\end{pmatrix} \, .
\end{align}
The relevant VMD Lagrangian is then~\cite{Fujiwara:1984mp}
\begin{align}
	\label{eq:LVMD}
	\cL_{\rm VMD}
=	\frac{g_{VVP}}{4}\langle  \boldsymbol{PV}^{\mu\nu}\boldsymbol{\tilde{V}}_{\mu\nu} \rangle
	-i g  \langle    \boldsymbol{V}^{\mu} ( \boldsymbol{P} \partial_\mu \boldsymbol{P}- \partial_\mu \boldsymbol{P P}) \rangle
	-  m^2_V \left( \frac{e}{g} \right) \langle \boldsymbol{V}_\mu \boldsymbol{Q} \rangle A^\mu \, ,
\end{align}
with $g_{VVP}= 3 g^2/(8\pi^2 f_\pi)$ and $g\approx\sqrt{12\pi} $.
Due to $a$--$P$ mixing, the first term in Eq.~\eqref{eq:LVMD} induces an $aVV$ vertex, the second term an $aVP$ vertex,
while the right-most term is the source of photon--vector-meson mixing.

The model considered in the Letter has $c_g\ne0$, and $c_q=c_{\gamma} = 0$.
Considering the Lagrangian of Eq.~\eqref{eq:LeffChiral}---at low masses, where this Lagrangian is valid---we obtain
\begin{align}
	\label{eq:M2aeta}
	M^2_{a\eta}
=	\frac{M^2_{a\eta'}}{2\sqrt{2}}
=	-\sqrt{\frac{2}{3}} B_0  \frac{m_d m_s m_u}{m_s m_d + m_s m_u + m_d m_u}
	\approx
	- \frac{m^2_{\pi^0}}{2\sqrt{6}}  \, ,
\end{align}
for the mass-mixing terms, where in the last step we take $m_s\gg m_d \approx 2m_u\,$, and
\begin{align}\label{eq: K explicit}
	K_{a\pi^0}
=&	\frac{1}{2}\frac{m_s(m_d-m_u)}{m_s m_u + m_d m_s + m_u m_d}
	\approx
	\frac{1}{6} \, ,  \nonumber \\
	K_{a\eta}
=&	\frac{1}{2} \sqrt{\frac{2}{3}}\frac{m_s(m_d+m_u)-m_u m_d}{m_s m_u + m_d m_s + m_u m_d}
	\approx
	\frac{1}{\sqrt{6}} \, ,  \\
	K_{a\eta'}
=&	\frac{1}{2} \sqrt{\frac{1}{3}}\frac{m_s(m_d+m_u) + 2m_u m_d}{m_s m_u + m_d m_s + m_u m_d}
	\approx
	\frac{1}{2\sqrt{3} } \,  . \nonumber
\end{align}
for the kinetic-mixing terms.
Therefore, the low-mass ALP $U(3)$ representation is given by Eq.~\eqref{eq:aSU3} with
\begin{align}
	\mixa{\pi^0}  &\approx \frac{\delta_I}{2} \frac{m_a^2}{m_a^2 - m_{\pi}^2}, \nonumber \\
  \mixa{\eta}  &\approx \left[ \frac{m_a^2}{\sqrt{6}} - \frac{m_{\pi^0}^2}{2\sqrt{6}}  \right] \frac{1}{m_a^2 - m_{\eta}^2}, \\
\mixa{\eta'}  &\approx \left[ \frac{m_a^2}{2\sqrt{3}}- \frac{m_{\pi^0}^2}{\sqrt{3}}  \right]  \frac{1}{m_a^2 - m_{\eta'}^2}, \nonumber
\end{align}
which gives the following values for the $\mathcal{C}_q$ terms:
\begin{align}
	2\sqrt{6}\cC_u
\approx &	\frac{m_a^2}{m_a^2 - m^2_{\pi^0}}
	+ \frac{2m_a^2 - m^2_{\pi^0}}{m_a^2 - m^2_{\eta}}
	+\frac{m_a^2 - 2m^2_{\pi^0}}{m_a^2 - m^2_{\eta'}}
	\, , \nonumber \\
	2\sqrt{6}\cC_d
\approx &	-\frac{m_a^2}{m_a^2 - m^2_{\pi^0}}
	+\frac{2m_a^2 - m^2_{\pi^0}}{m_a^2 - m^2_{\eta}}
	+\frac{m_a^2 - 2m^2_{\pi^0}}{m_a^2 - m^2_{\eta'}}
	 \, , \\
	2\sqrt{6}\cC_s
\approx&	-\frac{2m_a^2 -m^2_{\pi^0}}{m_a^2 - m^2_{\eta}}
	+2\frac{m_a^2 - 2m^2_{\pi^0}}{m_a^2 - m^2_{\eta'}} \nonumber
	 \, .
\end{align}
We note that at the limit of $m_a \gg m_{\eta'}$ these results give $\cC_s\to0$; however, the above equations are only valid for $m_a\lesssim 1\,$GeV.

\section{ALP decays}

\begin{figure}[]
  \includegraphics[width=0.49\textwidth]{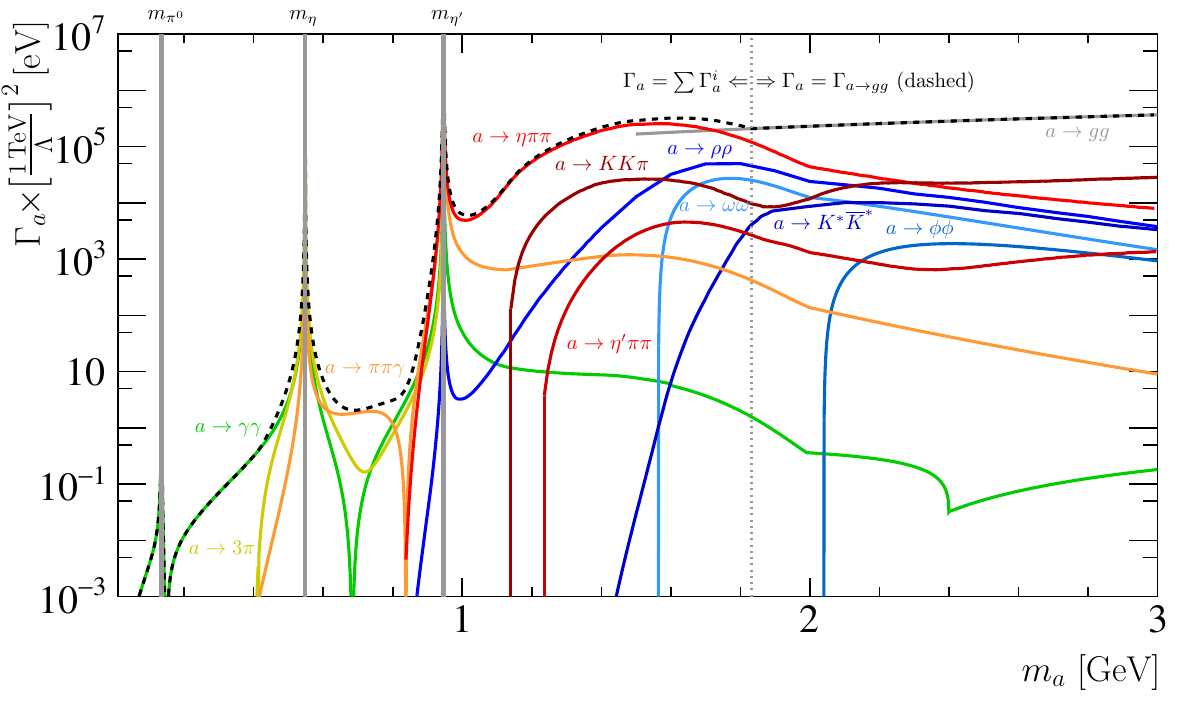}
  \includegraphics[width=0.49\textwidth]{a_bf.pdf}
\caption{ALP decay (left) widths and (right) branching fractions to all final states considered. For $m_a \lesssim 1.84$\,GeV, we take the total width to be the sum of the exclusive decay widths, whereas for $m_a \gtrsim 1.84$\,GeV we take the total width to be $\Gamma_{a\to gg}$.}
\label{fig:decay_rates}
\end{figure}

The decay rates and branching fractions are summarized in Fig.~\ref{fig:decay_rates}.
In this section, we provide the detailed calculations used to obtain these results.

\subsection{$a \to \gamma\gamma$}

Even though the ALP does not couple directly to the electromagnetic field when $c_{\gamma} = 0$, as shown in Eq.~\eqref{eq:c_eff} the chiral transformation generates a coupling at low masses.
In addition, ALP--pseudocalar mixing---followed by $P\to\gamma\gamma$---will also contribute.
Finally, at high masses and at the two-loop order, pQCD contributions from quarks become important.
The total decay rate for $a\to\gamma\gamma$ is given by
\begin{align}
	\label{eq:agammagamma}
	\Gamma_{a \to \gamma\gamma} = \frac{\alpha_{\rm EM}^2 m_a^3}{(4\pi)^3 f_a^2} \left| \mathcal{C}^{\chi}_{\gamma} +  \mathcal{C}_{\gamma}^{\rm VMD} + \mathcal{C}_{\gamma}^{{\rm pQCD},uds} +  \mathcal{C}_{\gamma}^{{\rm pQCD},cbt} \right|^2.
\end{align}
The contribution from the chiral transformation is
\begin{align}
	\label{eq:agammagamma_chiral_appendix}
	\mathcal{C}^{\chi}_{\gamma} = N_c \langle \boldsymbol{ \kappa QQ} \rangle \Theta(m_{\eta'}-m_a)  \approx \Theta(m_{\eta'}-m_a).
\end{align}
We turn this contribution off above the $\eta'$ mass, since the chiral rotation is no longer valid (see discussion in the main text on the $U(3)$ representation).
We calculate the VMD-based contribution as $a\to VV^{(\prime)}\to\gamma\gamma$, where the vector mesons mix with the photons, which predicts the pseudoscalar $P\to\gamma\gamma$ rates to $\mathcal{O}(10\%)$ accuracy.
This contribution is given by
\begin{align}
  	\mathcal{C}_{\gamma}^{\rm VMD}  &= -\mathcal{F}(m_a) \Theta(2.1\,{\GeV}-m_a) \left[ 3 \mixatwo{\rho}{\rho}
	+ \frac{1}{3} \mixatwo{\omega}{\omega}
	+ \frac{2}{3} \mixatwo{\phi}{\phi}
	+ 2  \mixatwo{\rho}{\omega} \right] \nonumber \\
&= 	-\mathcal{F}(m_a) \Theta(2.1\,{\GeV}-m_a) \frac{2 \tilde{\alpha}_s(m_a)}{3\sqrt{6}}\left(4\mathcal{C}_u + \mathcal{C}_d + \mathcal{C}_s  \right),
\end{align}
where the phenomenological suppression of the VMD amplitude at higher masses---obtained in the Letter using $e^+e^-$ data---is contained in the function $\mathcal{F}(m_a)$.
As we will show below, the pQCD-based contribution from light quarks surpasses the VMD-based one at $m_a \approx 2.1$\,GeV.
This is expected since, due to the suppression of the $VVP$ vertex at higher masses, contributions involving quark loops become dominant in the perturbative regime; therefore, we transition from the VMD-based light-quark contribution to the pQCD-based one at the point where the pQCD contribution is larger.
The full pQCD-based result has contributions from both light and heavy quarks\,\cite{Bauer:2017ris}
\begin{align}
	\mathcal{C}_{\gamma}^{{\rm pQCD},uds} &\approx \frac{\alpha_s^2(m_a)}{6 \pi^2} \left[ 5 \log{\frac{\Lambda^2}{m_{\pi}^2}}  + \log{\frac{\Lambda^2}{m_K^2}}  \right] \Theta(m_a - 2.1\,{\rm GeV}), \\
	\mathcal{C}_{\gamma}^{{\rm pQCD},cbt} &\approx  -\frac{\alpha_s^2(m_a) m_a^2}{72 \pi^2} \left[ \frac{4\sqrt{3}}{m_c^2}\log{\frac{\Lambda^2}{m_c^2}}  + \frac{1}{m_b^2}\log{\frac{\Lambda^2}{m_b^2}} + \frac{4}{m_t^2}\log{\frac{\Lambda^2}{m_t^2}}  \right] \Theta(m_a - 1.6\,{\rm GeV}).
\end{align}
These expressions are simplifications of those in Ref.~\cite{Bauer:2017ris}, and even though they are accurate to $\mathcal{O}(10\%)$ in the mass range that we use them, our numerical results are obtained using the full expressions.

\begin{figure}[t!]
\includegraphics[width=0.5\textwidth]{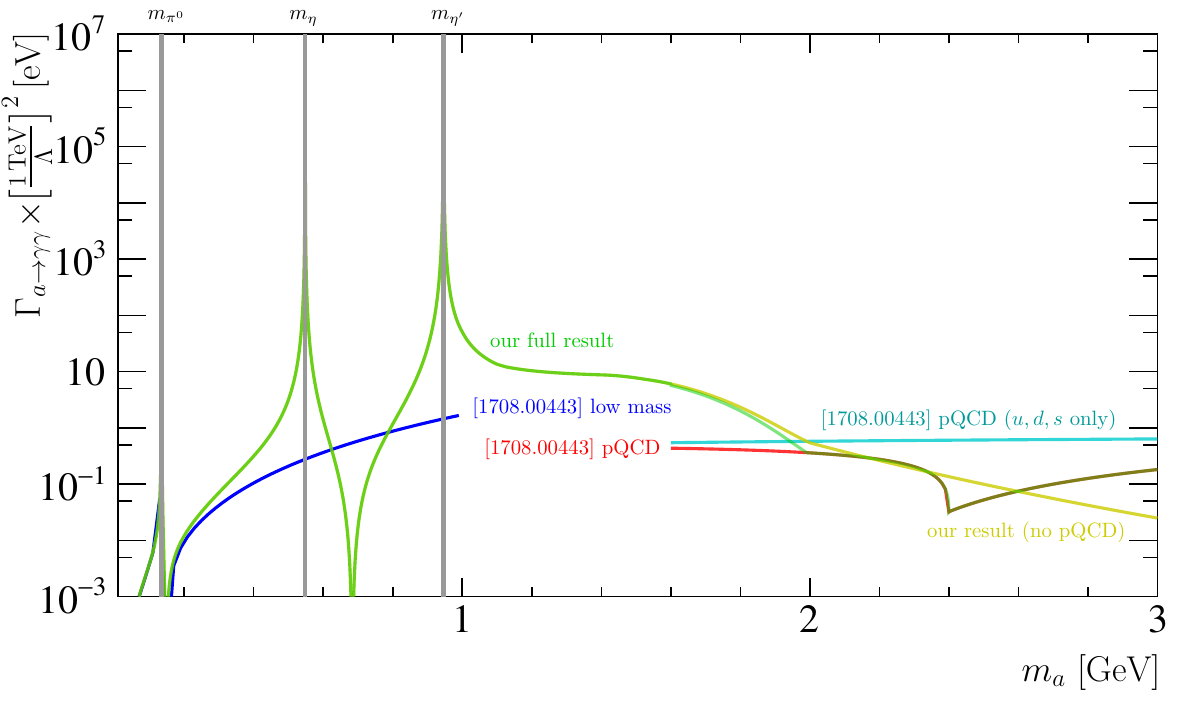}
\caption{The decay width for $a\to\gamma\gamma$ showing our result (green) with and (yellow) without the pQCD contribution, along with from Ref.~\cite{Bauer:2017ris} the (blue) low-mass and (red and cyan) pQCD contributions.}
\label{fig:agammagamma}
\end{figure}

Figure~\ref{fig:agammagamma} shows the various contributions to $\Gamma_{a\to\gamma\gamma}$ compared to those from Ref.~\cite{Bauer:2017ris}.
As expected, our result agrees with that of Ref.~\cite{Bauer:2017ris} for $m_a \lesssim 0.2$\,GeV and for $m_a \gtrsim 2.1$\,GeV, but is significantly different between these two mass regions.
This occurs because we include mixing with the $\eta$ and $\eta'$ mesons, and the VMD-based $a \to VV \to \gamma\gamma$ contribution.
Finally, one utility of the framework we are using is that one can immediately see that replacing the ALP by the pion, which includes neglecting the direct coupling to photons induced by the chiral transformation, gives the expected result:
\begin{align}
	\boldsymbol{a} \to {\boldsymbol \pi} \, , \
	m_a \to m_{\pi}\, ,\
	f_a \to f_{\pi}\, , \
	{\rm then}\, \nonumber \\
	\Gamma_{a \to \gamma\gamma} \to \frac{\alpha_{\rm EM}^2 m_{\pi}^3}{(4\pi)^3 f_{\pi}^2} = \Gamma_{\pi\to\gamma\gamma} \checkmark
\end{align}
The corresponding cross checks where the ALP is replaced by the $\eta^{(\prime)}$ also produce the well-known expected results.

\subsection{$a \to 3\pi$}

The decays $a \to 3\pi^0$ and $a\to\pi^+\pi^-\pi^0$ proceed via the isospin-violating $a$--$\pi$ mixing, and by $a$--$\eta^{(\prime)}$ mixing followed by $\eta^{(\prime)}\to3\pi$.
Given that these decays are explicitly isospin violating, we only calculate their rates up to $m_a = m_{\eta'}$ due to the large uncertainty in the isospin-violating component of the ALP $U(3)$ representation at higher masses.
Already by $m_{\eta'}$, these decays have small branching fractions.

Starting from the LO chiral Lagrangian, we find that $a\to 3\pi$ has contributions from the $4\pi$ and $2\eta_{(0,8)}2\pi$ vertices.
The former involves $a$--$\pi$ mixing, while the latter involves both $a$--$\eta_{(0,8)}$ and $\eta_{(0,8)}$--$\pi$ mixing.
The resulting amplitudes are
\begin{align}
	\label{eq:Aa3pi}
	 \mathcal{A}(a\to 3\pi^0)
=& 	\frac{m^2_\pi}{f_a f_\pi} \Bigg[ \mixa{\pi}
	- \delta_I\left( \frac{1}{\sqrt{3}} + \sqrt{2} S_{\eta\pi^0} +  S_{\eta'\pi^0} \right)
	\left( \sqrt{2} \mixa{\eta }+ \mixa{\eta^\prime }\right)  \nonumber\\
&	\qquad \qquad \qquad \qquad \qquad \qquad \qquad \qquad \qquad+ \sqrt{3}\delta_I \frac{m^2_\pi - 2m^2_\eta}{m^2_\pi - 4 m^2_\eta} \left(  \sqrt{2} S_{\eta\pi^0} + S_{\eta'\pi^0} \right)  \Bigg]  \,  ,\\
	\label{eq:Aapi0 pipls pimns}
	\mathcal{A}(a\to \pi^+\pi^-\pi^0)
=& 	\frac{1}{3f_a f_\pi} \Bigg\{
	(3 m_{\pi^+\pi^-}^2 - m_a^2 - 2m^2_\pi) \mixa{\pi}
	- \delta_I m^2_\pi \left( \frac{1}{\sqrt{3}} + \sqrt{2} S_{\eta\pi^0} +  S_{\eta'\pi^0} \right)
	\left(\sqrt{2} \mixa{\eta }+ \mixa{\eta^\prime }\right)  \nonumber\\
&	\qquad \qquad \qquad \qquad + \delta_I \frac{m^2_\pi - 2m^2_\eta}{m^2_\pi - 4 m^2_\eta}
	\left[\sqrt{3}m^2_\pi  \left(  \sqrt{2} S_{\eta\pi^0} + S_{\eta'\pi^0} \right) - 3m^2_{\pi^+\pi^-} + m^2_a + 3 m^2_\pi  \right]  \Bigg\}  \,  .
\end{align}
The decay rates are then
\begin{equation}
  	\Gamma_{a \to 3\pi}
= 	\frac{k}{2S m_a}\int |\mathcal{A}(a \to 3\pi) |^2 {\rm d}\Phi_3,
\end{equation}
where $S=1$ for $\pi^+\pi^-\pi^0$ and $S=3!$ for $3\pi^0$ are the usual symmetry factors.
The $k$ factor is added to account for the fact that the LO \chipt predictions for $\Gamma_{\eta^{(')}\to 3\pi}$ are a factor of $\approx 3$ lower than the corresponding experimental values.
The NNLO \chipt result is much larger than the LO calculation, largely due to final-state interactions between the pions\,\cite{Bijnens:2007pr}.
We use $k=2.7$ here, which is the mean of the $k$-factor values needed to obtain the known values of the $\eta$ and $\eta'$ decay widths, {\em i.e.}\ we obtain the $k$-factor by comparing to experimental data on $\eta^{(\prime)}\to 3\pi$ decays.
Given that the same $k$-factor works at $m_{\eta}$ and $m_{\eta'}$ to $\approx 20\%$ accuracy, we expect that this factor is reliable for $m_a \lesssim m_{\eta'}$; however, we have no reason to expect that this same $k$-factor works for higher masses, providing another motivation (beyond the large uncertainty on the isospin-violating component of the ALP $U(3)$ representation at higher masses discussed above) for only considering this decay below $m_{\eta'}$.

As above, we can again cross check our results by replacing the ALP with the low-mass pseudoscalars.
For example, using the same formalism we can derive the LO amplitude of $\eta_8\to 3\pi$
\begin{align}
	\Gamma_{a \to 3\pi^0}
& \to \frac{k}{12 m_{\eta}} \int \left| \frac{\delta_I m_{\pi}^2}{\sqrt{3} f_{\pi}^2} \right|^2  {\rm d}\Phi_3, \\
	{\rm and}\,\,
	\Gamma_{a \to \pi^+\pi^-\pi^0}
& \to \frac{k}{2 m_{\eta}} \int \left| \frac{\delta_I m_{\pi}^2}{\sqrt{3} f_{\pi}^2} \left[ \frac{ \frac{4}{3}m_{\pi}^2 - m_{\pi^+\pi^-}^2}{m_{\eta}^2 - m_{\pi}^2} \right] \right|^2  {\rm d}\Phi_3,
\end{align}
which are the well-known LO \chipt results for $\eta \to 3\pi$ when $k=1$.
As another check, considering only $a$--$\pi$ mixing gives the following:
\begin{align}
	 \mixa{\eta^{(\prime)}} &\to 0 \, , \ {\rm then} \nonumber\\
	 \Gamma_{a \to 3\pi}
&\to 	\frac{k m_a m_{\pi}^4}{32^2 \pi^3 f_a^2 f_{\pi}^2}\left| \delta_I
	\frac{m_a^2}{m_a^2 - m_{\pi}^2} \right|^2 \mathcal{K}_{3\pi} \left(\frac{m_{\pi}^2}{m_a^2} \right),
\end{align}
where
\begin{align}
	\mathcal{K}_{\pi^+\pi^-\pi^0}(x) & = \int_{4x}^{(1-\sqrt{x})^2} {\rm d}z \sqrt{1-\frac{4x}{z}} \left(x-z \right)^2 \sqrt{1 - 2(z+x) +(z-x)^2}, \\
	\mathcal{K}_{3\pi^0}(x) & = \frac{1}{3!} \int_{4x}^{(1-\sqrt{x})^2} {\rm d}z \sqrt{1-\frac{4x}{z}} \sqrt{1 - 2(z+x) +(z-x)^2}.
\end{align}
which agrees with Ref.~\cite{Bauer:2017ris} for $k=1$; {\em i.e.}\ our result agrees with that of Ref.~\cite{Bauer:2017ris}, except for our inclusion of ALP--$\eta^{(\prime)}$ mixing and the $k$ factor.

\subsection{$a \to VV$}

We calculate the decay rate for $a\to \rho\rho \to \pi_a \pi_b \pi_c \pi_d$ using VMD, including the phenomenological suppression factor obtained from $e^+e^-$ data. The amplitude is obtained from the $a\rho\rho$ vertex and is given by
\begin{equation}
  	\mathcal{A}(a \to 4\pi)
= 	\frac{3 g^4}{2 \pi^2 f_a} \varepsilon^{\mu\nu\alpha\beta} p_{\mu}^a p_{\nu}^b p_{\alpha}^c p_{\beta}^d
	\left[{\rm BW}_{\rho}(m_{ab}) {\rm BW}_{\rho}(m_{cd}) - {\rm BW}_{\rho}(m_{ad}) {\rm BW}_{\rho}(m_{bc}) \right]
	\mixa{\{\rho,\rho\}} \mathcal{F}(m_a) \, .
\end{equation}
 We use a mass-dependent width for the $\rho$ meson in the Breit-Wigner functions~(BW) following Ref.~\cite{Lees:2012cj}.
The $a \to 4\pi$ decay rates are then given by
\begin{equation}
  	\Gamma_{a \to 4\pi}
= 	\frac{1}{2S m_a}\int |\mathcal{A}(a \to 4\pi) |^2 {\rm d}\Phi_4 \, ,
\end{equation}
where $S=2$ for $\pi^+\pi^-\pi^0\pi^0$ and 4 for $2(\pi^+\pi^-)$ are the usual symmetry factors.
We can cross check this result by replacing the ALP with an $\eta'$ meson:
\begin{align}
	\boldsymbol{a} &\to {\boldsymbol \eta'} \, , \
	m_a \to m_{\eta'}, f_a \to f_{\pi} \ {\rm then} \nonumber\\
	\Gamma_{a \to 4\pi} &\to 58\,{\rm eV}
	\approx
	\Gamma_{\eta' \to 4\pi} = 52 \pm 13\,{\rm ev}~\checkmark
\end{align}
In the above cross check, we have summed the contributions from $a\to \pi^+\pi^-\pi^0\pi^0$ and $a \to 2(\pi^+\pi^-)$.
The experimental value is taken from Ref.~\cite{PDG}.

The decays $a \to \phi\phi \to 4K$ and $a\to K^*\bar{K}^* \to 2K2\pi$ are calculated in an identical way, but using the appropriate resonance parameters and symmetry factors.
Since the $\omega$ decays predominantly to $6\pi$, the Lorentz structure of the amplitude is more complicated.
Given that the $\omega$ is narrow, we instead calculate the decay rate of $a \to \omega\omega$ using the narrow-width approximation  and find
\begin{equation}
  	\Gamma_{a \to \omega\omega}
= 	\frac{9 m_a^3}{(4\pi)^5 f_a^2}\left|g^2 \mixatwo{\omega}{\omega} \mathcal{F}(m_a)  \right|^2
	\left(1 - \frac{4 m_{\omega}^2}{m_a^2}  \right)^{\frac{3}{2}} \, .
\end{equation}
 We do not consider the decay $a \to \phi\omega$ or any isospin-violating $VV$ decays.

We can now validate our data-driven approach by comparing the ALP branching fractions to the measured values of the corresponding $\eta_c$ branching fractions:
\begin{align}
	m_a \to m_{\eta_c},\,\,\text{Table~\ref{tab:VV} shows agreement for all}\,\,\eta_c \to VV\,\,\text{decays to}\,\mathcal{O}(10\%) \checkmark
\end{align}
The value of $f_a$ cancels in the branching fraction calculation, so it does not need to be specified in this comparison.
{\em N.b.}, the PDG does not quote a value for $\mathcal{B}(\eta_c \to \omega\omega)$ because no experiment has yet observed greater than $3\sigma$ evidence for this decay.
The value in Table~\ref{tab:VV} is the $\approx 2\sigma$ result from Ref.~\cite{Liu:2012eb}.
For the other three decays, the values in Table~\ref{tab:VV} are the PDG {\em average} values, {\em i.e.}\ the PDG averages of the experimental measurements.
In the main $\eta_c$ section of the PDG, the PDG instead quotes their {\em fit} values, which are the result of a constrained fit to a large number of $\eta_c$ decay observables.
We also note that since the $\eta_c$ can mix with $\eta^{(\prime)}$---and can decay electromagnetically---the ALP decay rates do not necessarily need to exactly match those of the $\eta_c$ meson, though we expect those effects to be small.
Finally, we note that mixing with the $\eta_c$ should also be considered for ALPs at this mass.
We leave this for future work.

The decay $a \to \pi\pi\gamma$ is calculated using $a\to\rho\rho$ followed by $\rho$--$\gamma$ mixing within the VMD framework. The result follows closely from those above and is given by
\begin{equation}
  	\Gamma_{a \to \gamma (\rho\to\pi\pi)}
= 	\frac{3 \alpha_{\rm EM} m_a^3}{2^{11} \pi^6 f_a^2}
	\int {\rm d}m_{\pi\pi}^2 \left| g^2 m_{\pi\pi} {\rm BW}_{\rho}(m_{\pi\pi}) \mixatwo{\rho}{\rho} \mathcal{F}(m_a)  \right|^2
	\left[ 1 - \frac{m^2_{\pi\pi}}{m_a^2}  \right]^3 \left[ 1 - \frac{4m_{\pi}^2}{m_{\pi\pi}^2} \right]^{\frac{3}{2}}  \, .
\end{equation}
This result is cross checked by comparing to the corresponding $\eta'$ decay:
\begin{align}
	\boldsymbol{a} &\to {\boldsymbol \eta'} \, , \ m_a \to m_{\eta'}\, , \ f_a \to f_{\pi} \ {\rm then} \nonumber\\
	\Gamma_{a \to \gamma \pi\pi} &\to 60\,{\rm keV} \approx \Gamma_{\eta' \to \pi\pi\gamma} = 56.6 \pm 1.0\,{\rm keV} \checkmark
\end{align}
Therefore, we also find the expected result for this decay, which is important for $m_a \lesssim 1$\,GeV.

\subsection{$a \to VP$}

\begin{figure}[]
\includegraphics[width=0.5\textwidth]{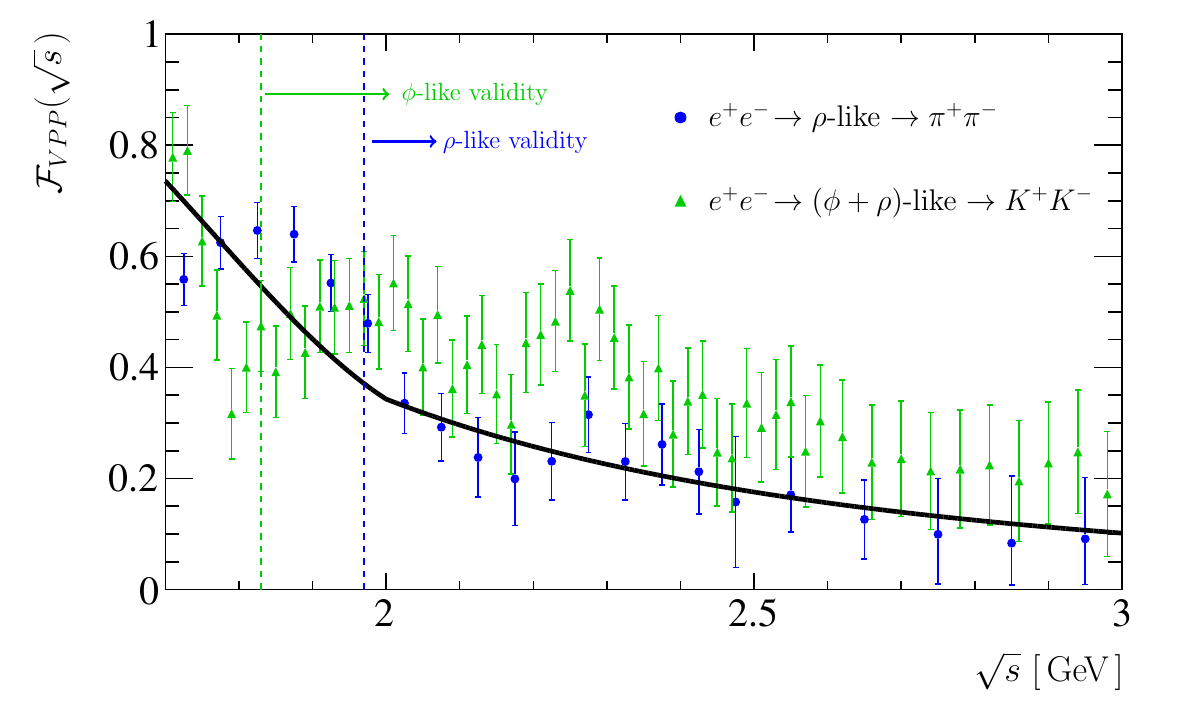}
\caption{Same as Fig.~\ref{fig:eedata} but for $e^+e^- \to V \to PP$ data\,\cite{Lees:2012cj,Lees:2013gzt}. The $\mathcal{F}$ function is the same as for the $VP$ final state, except for the pQCD scaling power due to the different dimensionality of the $VPP$ amplitude {\em c.f.}\ that of $VVP$. \label{fig:vpp}}
\end{figure}

Decays of the form $a \to VP$ proceed via the $VPP$ vertex.
As was done with $a \to VV$ in the Letter, the $a\to VP$ decay amplitude can be related to that of the $V \to P_1 P_2$ process via crossing symmetry, and this process can be studied using $e^+e^- \to V \to P_1 P_2$ data.
The amplitude for $V(p_V) \to P_1(p_1) P_2(p_2)$ is
\begin{align}
	\mathcal{A}(V \to P_1 P_2)
=  	g (p_1 - p_2)^{\mu}\epsilon_{\mu} \langle \boldsymbol{V}[\boldsymbol{P_1} ,\boldsymbol{P_2}] \rangle
	\mathcal{F}_{VPP}(p_V^2,p_1^2,p_2^2) \, ,
\end{align}
which is of a different dimension than the $VVP$ vertex for which $\mathcal{F}(m)$ was derived (this amplitude is one order lower in mass dimension).
Figure~\ref{fig:vpp} shows that $e^+e^- \to V \to P_1P_2$ data is consistent with using $\mathcal{F}_{VPP} = \mathcal{F}$, except with the pQCD scaling reduced by one order of mass dimension, {\em i.e.}\ the $[\beta_{\mathcal{F}}/m]^{4}$ term at high mass in Eq.~\eqref{eq:F} is replaced by $[\beta_{\mathcal{F}}/m]^{3}$ (the same constant $\beta_{\mathcal{F}}$ is used for both functions).

The decay rates $\Gamma_{a \to VP}$ can be calculated using VMD, along with the phenomenological suppression factor $\mathcal{F}_{VPP}(m)$ derived above.
Since $a \to \rho \pi$ violates isospin, we do not consider this mode.
The decay $a \to K^*K$ violates $U(3)$ symmetry for the case considered in the Letter, where $\mathcal{C}_u \approx \mathcal{C}_d \approx \mathcal{C}_s$; therefore, we will not explicitly calculate this decay rate here, though it is straightforward to do so given the equations in this subsection.
The absence of evidence for a resonance in $\eta_c \to K^*(\to K\pi)K$ supports our choice to neglect this channel.
Given that all other modes violate $C$, we do not consider any modes of this type.

\subsection{$a \to PPP$}

The amplitude for $a \to \eta\pi\pi$ includes contributions from a direct (mixing) term, corresponding to vertices such as $\eta_0\eta_8\pi\pi$ in the chiral Lagrangian, and from scalar and tensor resonances:
\begin{align}
	\mathcal{A}(a \to \eta\pi\pi)
= 	\frac{f_{\pi}}{f_a} \big[ \mathcal{A}_{\rm mix}(a \to \eta\pi\pi) +& \mathcal{A}(a \to \sigma(\pi\pi)\eta) \nonumber \\
	& + \mathcal{A}(a \to f_0(\pi\pi)\eta) + \mathcal{A}(a \to a_0(\eta\pi)\pi)  + \mathcal{A}(a \to f_2(\pi\pi)\eta) \big] \, .
\end{align}
The mixing term is given by
\begin{align}
	\mathcal{A}_{\rm mix}(a \to \eta\pi\pi)
	& \approx
	\left[ \sqrt{2} \mixa{\eta_0} + \mixa{\eta_8} \right] \frac{m_{\pi}^2}{3 f^2_{\pi}}   \mathcal{F}_{PPPP}(m_a) \approx 0 \, ,
\end{align}
where $\mathcal{F}_{PPPP}$ is the unknown $\mathcal{F}$ function for the 4-pseudoscalar vertex.
We include the mixing term, though its contribution is small for all masses, even for masses as low as $m_{\eta'}$ where the resonance contributions are all suppressed by their Breit-Wigner terms.

All of the necessary resonance parameters and couplings for the $\sigma$, $f_0$, and $a_0$ are taken from the $\eta' \to \eta\pi\pi$ model of Ref.~\cite{Fariborz:1999gr}, where the various resonance coupling constants were fit to all available data.
Accounting for ALP--pseudoscalar mixing gives the following terms for the $\pi^0\pi^0$ final state (the value of the amplitude for $\pi^+\pi^-$ is the same, though it involves different $U(3)$ generator expressions):
\begin{align}
	\mathcal{A}(a \to \sigma(\pi\pi)\eta)
& = 	- \left( \frac{10}{\rm GeV} \right)^2 \mixa{\eta\sigma}
	(p_a \cdot p_{\eta})(p_{\pi_1}\cdot p_{\pi_2}){\rm BW}_{\sigma}(m_{\pi_1\pi_2}) \mathcal{F}_{SPP}(m_a) \Theta(2 m_{K} -m_{\pi_1\pi_2})\, , \\
	\mathcal{A}(a \to f_0(\pi\pi)\eta)
& = 	\phantom{-} \left( \frac{7.3}{\rm GeV} \right)^2 \mixa{\eta f_0}
	(p_a \cdot p_{\eta})(p_{\pi_1}\cdot p_{\pi_2}){\rm BW}_{f_0}(m_{\pi_1\pi_2}) \mathcal{F}_{SPP}(m_a) \, , \\
	\mathcal{A}(a \to a_0(\eta\pi)\pi)
& = \phantom{-}	\left( \frac{13}{\rm GeV}\right)^2  \mixa{\pi^0 a_0} \mathcal{F}_{SPP}(m_a) \\
& 	\qquad \qquad \qquad \times \left[ (p_a \cdot p_{\pi_2})(p_{\eta}\cdot p_{\pi_1})  {\rm BW}_{f_0}(m_{\eta\pi_1})
	+ (p_a \cdot p_{\pi_1})(p_{\eta}\cdot p_{\pi_2}){\rm BW}_{f_0}(m_{\eta\pi_2}) \right]  \, , \nonumber
\end{align}
where the scalar-meson $U(3)$ representations were also fit to data and are approximately
\begin{align}
	\boldsymbol{\sigma} = \frac{1}{\sqrt{22}}{\rm diag}\{\sqrt{5},\sqrt{5},1\},\,
	\boldsymbol{f_0} = \frac{1}{2\sqrt{5}}{\rm diag}\{1,1,-2\sqrt{2}\}, \, \text{and}\,\,
	\boldsymbol{a_0} = \frac{1}{2}{\rm diag}\{1,-1,0\}.
\end{align}
We turn off the $\sigma$ contribution at the $KK$ threshold, since using a simple Breit-Wigner for this term above $2m_{K}$ violates unitarity.
An improved model could employ a coupled-channel K-matrix approach, though we do not consider this here.
We derive the $f_2$ amplitude and fix its couplings from its decay width to $\pi\pi$ and obtain
\begin{align}
	\mathcal{A}(a \to f_2(\pi\pi)\eta) = \left(\frac{16}{\rm GeV} \right)^2 \mixa{\eta f_2} & \left[  (p_{\eta} \cdot q_{\pi\pi})^2 - \frac{1}{3} q_{\pi\pi}^2 \left(p_{\eta} - p_{\pi\pi} \left( \frac{p_{\eta} \cdot p_{\pi\pi}}{p_{\pi\pi}^2} \right) \right)^2 \right]
	{\rm BW}_{f_2}(m_{\pi_1\pi_2}) \mathcal{F}_{TPP}(m_a) \, ,
\end{align}
where $p_{\pi\pi} \equiv p_{\pi_1} + p_{\pi_2}$, $q_{\pi\pi}  \equiv p_{\pi_1} - p_{\pi_2}$, and for simplicity we take $\boldsymbol{f_2} = {\rm diag}\{1,1,0\}/2$, which is known to be a good approximation~\cite{PDG}.
The expressions for $a \to \eta'\pi\pi$ are similar, though with $\boldsymbol{\eta}$ replaced by $\boldsymbol{\eta'}$ and the $a_0 \to \eta'\pi$ coupling coupling constant is 20\% larger than that of $a_0 \to \eta\pi$.

Unlike for the $VVP$ and $VPP$ vertices, we cannot derive the $\mathcal{F}_{PPPP}$, $\mathcal{F}_{SPP}$ and $\mathcal{F}_{TPP}$ functions from $e^+e^-$ data.
At low masses, these $\mathcal{F}$ functions are normalized to be unity just like the $VVP$ and $VPP$ ones.
Furthermore, the pQCD power counting at high masses is the same for all of these amplitudes as it is for $VVP$, assuming that the tetraquark content of all resonances is small.
Therefore, we take
\begin{align}
	\mathcal{F}_{SPP}(m) = \mathcal{F}_{PPPP}(m) = \mathcal{F}_{TPP}(m) = \mathcal{F}(m) \, .
\end{align}
This assumption is shown below to have better than $\mathcal{O}(1)$ accuracy, though we note here that an improved understanding of the excited $\eta^*$ states would enable deriving better data-driven constraints on $\mathcal{F}_{SPP}$, $\mathcal{F}_{TPP}$, and $\mathcal{F}_{PPPP}$.

Using the amplitudes defined above, the rates are obtained as
\begin{equation}
  	\Gamma_{a \to \eta^{(\prime)}\pi\pi}
= 	\frac{1}{2 S m_a}\int |\mathcal{A}(a \to \eta^{(\prime)}\pi\pi) |^2 {\rm d}\Phi_3 \, ,
\end{equation}
where $S=2$ for $\eta^{(\prime)}\pi^0\pi^0$ and 1 for $\eta^{(\prime)}\pi^+\pi^-$ are the usual symmetry factors.
First, we cross check our result for the $a\to \eta\pi\pi$ decay by replacing the ALP with an $\eta'$, which gives (summing the $\eta\pi^+\pi^-$ and $\eta\pi^0\pi^0$ modes)
\begin{align}
	\boldsymbol{ a} &\to {\boldsymbol \eta'} \, ,
	m_a \to m_{\eta'}\, ,
	f_a \to f_{\pi} \ {\rm then} \nonumber\\
	\Gamma_{a \to \eta\pi\pi} &\to 116\,{\rm keV} \approx \Gamma_{\eta' \to \eta\pi\pi} = 128 \pm 2\,{\rm keV} \checkmark
\end{align}
Of course, since the model of Ref.~\cite{Fariborz:1999gr} was fit to data---including this $\eta'$ decay---the numerical value should be similar if implemented correctly.

A more interesting cross check involves comparing the ALP branching fractions to the corresponding known $\eta_c$ values for $m_a = m_{\eta_c}$.
We first perform this check for $\eta_c \to \eta\pi\pi$:
\begin{align}
	m_a \to m_{\eta_c}, \,\text{then}\,\, \mathcal{B}(a \to \eta\pi^+\pi^-) \to 1.5\% \approx \mathcal{B}_{\rm PDG}(\eta_c \to \eta\pi^+\pi^-) =  1.7\pm0.5\% \checkmark
\end{align}
Performing the same check for $a\to\eta'\pi^+\pi^-$ gives a prediction of $0.3\%$, whereas the experimental value from Belle is $1.3\pm 0.2\%$~\cite{Xu:2018uye}. 
However, Belle attributes $\approx 64\%$ of $\eta_c \to \eta'\pi\pi$ to a 2\,GeV scalar resonance that is not included in our model.
(After submitting this Letter to arxiv, Belle updated their paper to remove the claim about the 2\,GeV scalar resonance. However, it is clear that much of this decay involves high-mass dipions whose source is not included in our model.)
Our prediction is consistent with the remaining branching fraction of $\approx 0.5\pm0.1\%$.
Furthermore, our predictions for $\eta_c \to \eta' f_0(980)$ and $\eta_c \to \eta' f_2(1270)$ are both consistent with the published dipion mass spectrum in Ref.~\cite{Xu:2018uye}. 
So, while we underestimate the $\eta'\pi\pi$ branching fraction for $m_a$ values close to 3\,GeV, for lower masses our prediction for this final state should be within a factor of two (we do not see any need to improve the prediction for this final state).
Therefore, we conclude that our predictions for $a \to \eta^{(\prime)}\pi\pi$ are consistent with $\eta_c$ data, which validates our $\mathcal{F}_{SPP}$, $\mathcal{F}_{PPPP}$, and $\mathcal{F}_{TPP}$ functions with better than $\mathcal{O}(1)$ accuracy.

Finally, there is one additional cross check that can be performed using the $\eta(1760)$ state:
\begin{align}
	m_a &\to m_{\eta(1760)}\, , \ \text{then} \nonumber\\
	\mathcal{B}(a\to\gamma\gamma)\times  \mathcal{B}(a\to\eta'\pi^+\pi^-)\to 1.1 \cdot 10^{-7}
	&\approx \mathcal{B}(\eta(1760)\to\gamma\gamma)\times\mathcal{B}(\eta(1760)\to\eta'\pi^+\pi^-) = (1.2 \pm 0.3) \cdot 10^{-7}\checkmark
\end{align}
Here, we have assumed that the $U(3)$ representation of the $\eta(1760)$ is the same as that of the ALP.
Lack of knowledge of the nature of this $\eta(1760)$ state induces an $\mathcal{O}(1)$ uncertainty here.
Taken together, we conclude that the available cross checks suggest that our $\mathcal{F}_{SPP}$, $\mathcal{F}_{PPPP}$, and $\mathcal{F}_{TPP}$ functions are accurate with at most $\mathcal{O}(1)$ uncertainty over the full mass range considered in this study.

The rate for the family of decays $a \to K\overline{K}\pi$ is calculated using a similar approach to the one used above for $a \to \eta\pi\pi$.
We take the total amplitude to be
\begin{align}
	\mathcal{A}(a \to K\overline{K}\pi)
= 	\frac{f_{\pi}}{f_a} [ \mathcal{A}(a \to S_{K\pi}(K\pi)K) + \mathcal{A}(a \to a_0(KK)\pi) ] \, ,
\end{align}
where $S_{K\pi}$ denotes the $K\pi$ $S$-wave amplitude.
It is well-known that $S_{K\pi}$ has a large $K^*_0(1430)$ contribution, and that it is not well described by a simple sum of Breit-Wigner terms.
We use the empirical $S_{K\pi}$ amplitude measured in Ref.~\cite{Lees:2015zzr} by {\sc BaBar}.
In principle, there is a mixing term similar to the $a \to \eta\pi\pi$ case; however, it is negligible for all $m_a$ and so we ignore it.

Here we provide the full expressions for the amplitudes for the $a \to K^+K^-\pi^0$, though as above the value of the amplitude is the same for all 6 $K\overline{K}\pi$ final states (but involves different $U(3)$ generator expressions):
\begin{align}
	\mathcal{A}(a \to a_0(K^+K^-)\pi^0) & =  \left( \frac{13}{\rm GeV} \right)^2 \mixa{a_0 \pi^0}
		(p_a \cdot p_{\pi})(p_{K^+}\cdot p_{K^-}){\rm BW}_{a_0}(m_{KK}) \mathcal{F}_{SPP}(m_a) \, , \\
	\mathcal{A}(a \to S_{K\pi}(K\pi)K) & = \left( \frac{8.2}{\rm GeV}\right)^2 \mixa{\{ K^+,K^-\}} \mathcal{F}_{SPP}(m_a)  \\
	& \qquad \qquad \qquad \times \left[ (p_a \cdot p_{K^-})(p_{\pi}\cdot p_{K^+}) S_{K\pi}(m_{K^+\pi}) +  (p_a \cdot p_{K^+})(p_{\pi}\cdot p_{K^-}) S_{K\pi}(m_{K^-\pi}) \right]  \, , \nonumber
\end{align}
where again the coupling parameters are taken from Ref.~\cite{Fariborz:1999gr} (the $\kappa$ couplings are used for $S_{K\pi}$), and as stated above, the shape and phase of $S_{K\pi}$ is taken from Ref.~\cite{Lees:2015zzr}.
The total rate for $a \to K\overline{K}\pi$ is 6 times that obtained for $a\to K^+K^-\pi^0$ due to equivalent contributions also from $K^{\pm}K_S\pi^{\mp}$, $K^{\pm}K_L\pi^{\mp}$, and $K_SK_L\pi^0$.

We cross check our $a\to K\overline{K}\pi$ result by comparing to the corresponding known $\eta_c$ values for $m_a = m_{\eta_c}$:
\begin{align}
	m_a \to m_{\eta_c}, \,\text{then}\,\, \mathcal{B}(a \to K\overline{K}\pi) \to 7.8\% \approx \mathcal{B}_{\rm PDG}(\eta_c \to K\overline{K}\pi) =  7.3\pm0.5\% \checkmark
\end{align}
Our calculation is consistent with the measured value.
{\em N.b.}, while the shape and phase of $S_{K\pi}$ were taken from a fit to data\,\cite{Lees:2015zzr}, the magnitude is set by the $\kappa$ resonance parameters in Ref.~\cite{Fariborz:1999gr} and by our $\mathcal{F}$ function.

\section{ALP Constraints}

Here we provide details on the constraints placed on the ALP scenario discussed in the Letter, {\it i.e.} the case where only $c_g\ne0\,$.
We focus on the mass region of $m_\pi < m_a < 3\,$GeV, but note that for $m_a<100\,$MeV the strongest constraint is from $\BR(K^+ \to \pi^+ + {\rm invisible})<7.3 \times 10^{-11}$~\cite{Artamonov:2009sz}, where $f_a \gtrsim 3\,$TeV~\cite{Fukuda:2015ana,Bauer:2018uxu}.
Constraints where $f_a \lesssim 3f_{\pi}$ are omitted, {\em e.g.}, bounds from radiative $J/\psi$ decays, since we assumed $f_{\pi} \ll f_a$ when deriving the ALP--pseudoscalar mixing factors.

\subsection{LEP \& Beam Dumps}

Limits have been placed on the $a\gamma\gamma$ vertex from LEP\,\cite{Abbiendi:2002je,Knapen:2016moh} and beam-dump experiments\,\cite{Bjorken:1988as,Blumlein:1990ay}.
We use the $a\to\gamma\gamma$ calculation above to relate the $a\gamma\gamma$ interaction strength to $f_a$.
The beam-dump limits only constrain ALP masses where $\mathcal{B}(a \to \gamma\gamma) = 1$, even for the scenario considered here; therefore, the relationship between the ALP lifetime and the strength of the $a\gamma\gamma$ vertex is the same here as it is when the ALP only interacts with the electromagnetic field.
For the LEP constraints, we also include our calculation of $\mathcal{B}(a \to \gamma\gamma)$ when recasting the published limits for this model.

\subsection{$\phi\to a\gamma$}

Constraints can be placed on $f_a$ considering the decays $\phi \to a\gamma$, with $a \to \pi\pi\gamma$ and $a\to\eta\pi^0\pi^0$,
using experimental upper limits for the $\phi\to \pi\pi \gamma \gamma$ and $\phi \to \eta\pi^0\pi^0\gamma$ decays\,\cite{PDG}:
\begin{align}
	\mathcal{B}(\phi \to \pi\pi\gamma\gamma) < 1.2 \times 10^{-4}\, , \qquad
	\mathcal{B}(\phi \to \eta\pi^0\pi^0\gamma) < 2 \times 10^{-5}\, ,
\end{align}
where in each case we conservatively assume that the ALP decay is the only contribution to each final state.
We then use the ratio of $\phi$ decay rates
\begin{align}
	\frac{\Gamma_{\phi \to a\gamma}}{\Gamma_{\phi\to\eta\gamma}}
= 	\left[ \frac{f_{\pi} \langle\boldsymbol{a\phi\phi}\rangle }{f_a  \langle\boldsymbol{\eta\phi\phi}\rangle } \right]^2
	\left[ \frac{m_{\phi}^2 - m_a^2}{m_{\phi}^2 - m_{\eta}^2} \right]^3 \, ,
\end{align}
the known value $\mathcal{B}(\phi \to \eta\gamma) = 1.3\%$\,\cite{PDG}, and the ALP branching fractions shown in Fig.~\ref{fig:decays} to determine the constraints on $f_a$.

\subsection{$\eta' \to a\pi\pi$}

The decay $\eta' \to \pi^+\pi^-a$ with $a\to\pi^+\pi^-\pi^0$ is used to place constraints on $f_a$ using the experimental upper limit\,\cite{PDG}
\begin{align}
	\mathcal{B}(\eta' \to 2(\pi^+\pi^-)\pi^0) < 1.8 \times 10^{-3}.
\end{align}
We calculate the rate for $\eta' \to \pi^+\pi^-a$ using our $a \to \eta\pi\pi$ model, but with the parent particle properties replaced by those of the $\eta'$ and the final-state $\eta$ replaced by the ALP:
\begin{align}
	\mathcal{A}(\eta' \to \pi\pi a) = \mathcal{A}(a \to \eta\pi\pi)\{ m_a \to m_{\eta'}, m_{\eta} \to m_a, \mixa{\eta'} \to 1, f_a \to f_{\pi} \} \times \mixa{\eta}
\end{align}
{\em i.e.}, we take the amplitude for $a\to\eta\pi\pi$ but for $m_a = m_{\eta'}$, $f_a = f_{\pi}$, and $\boldsymbol{a} = \boldsymbol{\eta'}$, along with also $m_{\eta} = m_a$, then multiply the result by $\mixa{\eta}$ which accounts for the ALP-$\eta$ mixing.
We then normalize this using the known value of $\mathcal{B}(\eta' \to \eta\pi\pi)$\,\cite{PDG}.
Using our calculation of $\mathcal{B}(a \to \pi^+\pi^-\pi^0)$ we are able to determine the constraints on $f_a$.

\subsection{Penguin Decays}

First, we consider $b \to s a$ production in penguin decays.
At one loop, the $agg$ vertex generates an axial-vector $a tt$ coupling~\cite{Bauer:2017ris} which results in the following exclusive decay branching fractions~\cite{Batell:2009jf,Hiller:2004ii,Bobeth:2001sq,Choi:2017gpf}:
\begin{align}
	\label{eq:btosbf}
	\mathcal{B}(B \to K a) & \approx 0.03 \left[ \frac{10\,{\rm TeV} c_g \alpha_s^2(m_t) \mathcal{UV}}{\Lambda} \right]^2 \left[\frac{1}{1 - m_a^2 / 38\,{\rm GeV}^2}  \right]^2 \lambda^{\frac{1}{2}}(m_a,m_K), \\
	\mathcal{B}(B \to K^* a) & \approx  0.04 \left[ \frac{10\,{\rm TeV} c_g \alpha_s^2(m_t) \mathcal{UV}}{\Lambda} \right]^2 \left[\frac{3.65}{1 - m_a^2 / 28\,{\rm GeV}^2}  - \frac{2.65}{1 - m_a^2 / 37\,{\rm GeV}^2} \right]^2 \lambda^{\frac{3}{2}}(m_a,m_{K^{*}}),
\end{align}
where the usual kinematic factors are
\begin{align}
	\lambda(m_a,m_{K^{(*)}}) = \left[1 - \left(\frac{m_a + m_{K^{(*)}}}{m_B}\right)^2\right]\left[1 - \left(\frac{m_a - m_{K^{(*)}}}{m_B}\right)^2\right].
\end{align}
The loop contains a factor that depends on the UV physics and is schematically given by
\begin{align}
	\mathcal{UV} \approx \log{\frac{\Lambda^2_{\rm UV}}{m^2_t}} \pm \mathcal{O}(1) \Rightarrow 1,
\end{align}
which we simply take to be unity given that the log factor is also $\mathcal{O}(1)$ for the scales being probed by currently available data.
See Ref.~\cite{Freytsis:2009ct} for detailed discussion on UV completion in this context, where an explicit UV completed model is presented.
Clearly this choice of UV factor induces $\mathcal{O}(1)$ arbitrariness on the constraints placed on ALPs in $b\to s$ penguin decays.

Using Eqs.~\eqref{eq:btosbf} and the ALP decay branching fractions calculated in the Letter, we use the following experimental data to constrain $\Lambda$:
\begin{itemize}
	\item Ref.~\cite{Aubert:2008bk} reported $\mathcal{B}(B^{\pm} \to K^{\pm} \eta(1295))\times\mathcal{B}(\eta(1295)\to\eta\pi^+\pi^-) = 2.9 \pm 0.8 \times 10^{-6}$. Ref.~\cite{Aubert:2008bk} also provides the $m_{\eta\pi\pi}$ spectrum up to 1.5\,GeV, from which it is clear that the $\eta(1295)$ is the largest peaking signal in the spectrum, other than the $\eta'$ (the $\eta(1295)$ is much broader than the detector resolution).
	Based on this, we take $\mathcal{B}(B^{\pm} \to K^{\pm} a) \times \mathcal{B}(a \to \eta \pi^+\pi^-) < 2 \times 10^{-6}$ for $m_a < 1.5$\,GeV, excluding the $\eta'$ peak region.
	\item Ref.~\cite{Aubert:2008bk} also reported $\mathcal{B}(B^{\pm} \to K^{\pm} \eta(1475))\times\mathcal{B}(\eta(1475)\to K^*K) = 1.4 \pm 0.2 \times 10^{-5}$ using the $K^{\pm}K_S\pi^{\mp}$ final state, where at least one $K\pi$ combination was required to be in the window $0.85 < m_{K\pi} < 0.95$\,GeV.
	While the decay $a \to K^*K$ violates $SU(3)$ symmetry in the model considered here, $a \to K\overline{K}\pi$ decays often produce a $K\pi$ pair that falls within this $K^*$ mass window;
	we numerically calculate $\mathcal{B}(a \to KK\pi)$ for ${0.85<m_{K\pi}<0.95\,{\rm GeV}}$.
	Via inspection of the published $K^*K$ mass spectrum, which is shown up to 1.8\,GeV, we take
	$\mathcal{B}(B^{\pm} \to K^{\pm} a) \times \mathcal{B}(a \to K^{\pm}K_S\pi^{\mp}) < 1 \times 10^{-7}$ for $0.85<m_{K\pi}<0.95\,{\rm GeV}$ and $m_a < 1.8$\,GeV.
	\item Using $\mathcal{B}(B^0 \to K^0 \phi\phi) = 4.5 \pm 0.9 \times 10^{-6}$\,\cite{Lees:2011zh} we assume that the entire decay rate is due to ALPs and take $\mathcal{B}(B^0 \to K^0 a) \times \mathcal{B}(a \to \phi \phi) < 6 \times 10^{-6}$.
	\item Using $\mathcal{B}(B^{\pm} \to K^{\pm} \omega(3\pi)) = 5.9 \pm 0.4 \times 10^{-6}$\,\cite{Chobanova:2013ddr} we assume that the entire decay rate is due to ALPs and take $\mathcal{B}(B^{\pm} \to K^{\pm} a) \times \mathcal{B}(a \to \pi^+\pi^-\pi^0) < 6.5 \times 10^{-6}$ for $0.73 < m_a < 0.83$\,GeV. From inspection of the $3\pi$ mass plot in Ref.~\cite{Chobanova:2013ddr}, which only shows this mass range, we conclude that taking the limit to be constant in this region is conservative.
\end{itemize}
Finally, we also constrain $\Lambda$ by setting an upper limit on the inclusive $b \to s a$ rate. Since the ALPs considered here are not massive enough to decay into charm hadrons, the observed inclusive $b \to c$ branching fractions require that
\begin{align}
	\mathcal{B}(b \to s a) \approx 5 \times \left[ \mathcal{B}(B \to K a) + \mathcal{B}(B \to K^*a) \right] < \left[1 - \mathcal{B}(b \to c)\right] \lesssim 5\%,
\end{align}
where the factor of 5 is consistent with the known value of the ratio of $\mathcal{B}(b \to s \mu\mu)$ and $[ \mathcal{B}(B \to K \mu\mu) + \mathcal{B}(B \to K^*\mu\mu) ]$\,\cite{PDG}.

Similarly, $s \to d$ penguin decays can also be used to place constraints on $\Lambda$.
Ref.~\cite{Izaguirre:2016dfi} used existing data on $K^{\pm} \to \pi^{\pm}\gamma\gamma$\,\cite{Ceccucci:2014oza} and $K_L \to \pi^{0}\gamma\gamma$\,\cite{Abouzaid:2008xm} to place constraints on ALPs that dominantly couple to electroweak bosons.
We recast these limits for the ALP model considered here.
Note that $\mathcal{B}(a\to\gamma\gamma)\approx 1$ for the ALP masses probed in these decays, even when the dominant coupling is to gluons.

\subsection{Additional Constraints}

We leave for future work detailed studies of other production mechanisms.
For example, 
higher-mass ALP production via $\gamma\gamma\to a$ fusion, {\em e.g.}\ in ultra-peripheral PbPb collisions\,\cite{Knapen:2016moh}, can be replaced by central exclusive production at the LHC, {\em etc.}

\end{document}